\documentclass[]{aa}  

\usepackage{graphicx}
\usepackage{txfonts}
\usepackage{orcidlink}
\usepackage{subcaption}         
\usepackage{lscape}             
\usepackage{placeins}           
\usepackage{natbib}
\bibpunct{(}{)}{;}{a}{}{,}  

\usepackage{soul} 

\usepackage[]{hyperref}
\begin{document}
\title{Thermal instability in coronal loops: linking eigenvalue spectra to time-dependent evolution}
\titlerunning{Thermal instability in coronal loops}
\authorrunning{Kelly et al.}

\author{Adrian Kelly\inst{}\thanks{Corresponding author: \texttt{adrian.kelly@kuleuven.be}} \orcidlink{0009-0006-8524-008X}
  \and Rony Keppens\inst{} \orcidlink{0000-0003-3544-2733}
  \and Jordi De Jonghe\inst{} \orcidlink{0000-0003-2443-3903}}
  
\institute{Centre for mathematical Plasma-Astrophysics (CmPA), KU Leuven, Celestijnenlaan 200B, 3001 Leuven, Belgium}

\date{}

\abstract
{Cool, dense condensations such as coronal rain and prominences suggest that coronal plasma can undergo runaway radiative cooling. Connecting this behaviour to linear thermal modes requires us to fully understand the deeper connection between eigenvalue spectra and actual time-dependent evolution.}
{We aim to clarify this intricate link for a simplified, coronal-only model of a stratified coronal loop by combining spectral, linear initial-value, and nonlinear simulations of the same loop setup.}
{We study waves and instabilities, as well as temporal evolutions for a 1D hydrostatic, thermally balanced loop with optically thin radiation and prescribed heating. The non-adiabatic spectrum of all the physically realisable eigenmodes is computed with our open-source code \textsc{Legolas}. We demonstrate here our newly developed boundary value-initial value solver called \textsc{Legolas-IVP}, where linear evolutions are performed for controlled perturbations, and fully equivalent nonlinear runs are carried out with our generic software toolkit \textsc{MPI-AMRVAC}.}
{The spectrum of the stratified 1D loop contains discrete acoustic modes and a thermally unstable branch consisting of thermal modes, including a thermal continuum. Linear initial-value experiments with isochoric, isobaric, and isentropic pulses highlight how the polarisation of the eigenmodes and the obtained evolution from specific perturbations demonstrate physically consistent behaviour expected from the eigenspectrum. Even in the linear stage, thermal imbalance drives siphon-like flows from the footpoints toward the cooling region. Growth rates measured from \textsc{Legolas-IVP} agree with the spectral predictions and are reproduced in \textsc{MPI-AMRVAC}. The latter follows the condensation through runaway cooling to chromospheric temperatures, with the resulting cool dense blob sliding under gravity toward the loop footpoint.}
{The spectral--linear--nonlinear investigation for the simple 1D loop demonstrates a direct link between thermal eigenmodes and time-dependent condensation dynamics, and provides a basis for extending such mode-based interpretations to fully 3D magnetohydrodynamic models.}

               
\maketitle 
\nolinenumbers

\section{Introduction}
\subsection{Context and motivation}
Coronal rain and prominences are striking examples of cool, dense condensations forming within the million–degree solar corona. Our current theoretical understanding and the sustained progress in modelling efforts for forming these enigmatic condensations is reviewed in \citet{RK2025,liakh_numerical_2025,yuhao2025}.
Hot coronal plasma in loops is continuously losing energy by radiation and thermal conduction and can be susceptible to localised cooling instabilities, even in the presence of strong background heating.
Thermal instability (TI) \citep{parker_instability_1953, field_thermal_1965} is widely thought to drive the in-situ formation of these condensations by triggering runaway cooling that is dictated by the precise density-temperature dependence of the radiative loss function. 
As coronal loops are gravitationally and thermally stratified, and because their coupling to denser chromospheric regions can regulate thermal non-equilibrium (TNE) \citep{antiochos_model_1991}, both TI and TNE are thought to play crucial roles in driving cyclic heating-cooling behaviour in coronal loops \citep{muller_dynamics_2003}.
This may explain observed intensity pulsations and recurrent coronal rain \citep{froment_occurrence_2018}.

While TNE is tied to the chromosphere-transition region–corona stratification of the solar atmosphere, TI is not specific to the solar corona. Rather, it is a general mechanism invoked to explain non-gravitationally driven condensations in optically thin, radiatively cooling plasmas.
Particularly, in the interstellar medium (ISM) and intracluster medium (ICM), thermal instability has been used to explain the coexistence of hot, warm, and cold gas phases \citep{field_cosmic-ray_1969, lepp_thermal_1985}. 
TI-driven radiative turbulent mixing layers have been identified as a key site of mass and energy exchange in multiphase ICM plasma \citep{fielding_multiphase_2020}, and cloud--wind interactions in galactic outflows have been shown to drive in situ cold gas formation through combined shear and radiative cooling \citep{banda-barragan_shockmulticloud_2021}. 

Despite decades of work, many fundamental aspects of coronal condensation processes are still debated.
Open questions include the precise relation between TI and TNE \citep{klimchuk_distinction_2019}, the role of background heating and its detailed spatio-temporal variation \citep{Brughmans2022,yoshihisa_conditions_2025}, and the extent to which observed condensations can be interpreted as the nonlinear outcome of linear thermal eigenmodes. 
Regarding the latter, fully nonlinear magnetohydrodynamic (MHD) simulations have successfully reproduced prominence formation \citep{donne_mass_2024}, coronal rain \citep{Fang2015}, and cyclic evaporation–condensation behaviour \citep{zekun2024}, but the degree of realism achieved in these multi-dimensional MHD models makes it challenging to clearly identify whether particular linear instabilities control the early evolution, and which physical perturbations are responsible for the observed dynamics.
Bridging this gap between linear theory and fully nonlinear simulations is essential to interpret condensations as the natural outcome of the underlying (stable and unstable) eigenmodes of the system, and to connect these insights to other multiphase environments.

We consider a 1D coronal loop model, where we isolate the coronal part alone, deliberately excluding the important processes of thermal conduction and coupling to the chromospheric regions. This simplification results in a well-defined, easily reproducible and controlled setting in which to study how TI and runaway cooling operate in gravitationally-stratified atmospheres.
A combination of a linear spectral code and a nonlinear (M)HD code are used, which operate on the same equilibrium. Our 1D loop is a stratified, semi-circular shape as typically observed in the solar corona, and frequently adopted in 1D models.

For the linear analysis we use the \textsc{Legolas} code, an eigenvalue solver designed to compute full non-adiabatic (M)HD spectra including thermal effects \citep{claes_legolas_2020, claes_legolas_2023}, which employs a suitable finite element spatial discretisation that allows the representation of both discrete global modes, as well as a dense but discrete subset of the ultra-localised continuum modes. 
This paper also serves to introduce and demonstrate a newly developed initial-value solver called \textsc{Legolas-IVP}, built on the same discretisation.
For the nonlinear regime we employ \textsc{MPI-AMRVAC} \citep{xia_mpi-amrvac_2018, keppens_mpi-amrvac_2021, keppens_mpi-amrvac_2023}, a parallelised simulation code for solving multi-dimensional partial differential equations that allows us to follow the evolution into the fully nonlinear phase. While \textsc{MPI-AMRVAC} can be employed in any dimensionality \citep{liakh_numerical_2025}, we here use it for 1D non-adiabatic hydrodynamic evolutions for coronal loops.

\subsection{(Magneto)hydrodynamic spectroscopy}
MHD spectral theory \citep{goedbloed_mhd_1993, goedbloed_magnetohydrodynamics_2019} shows that the linear spectrum of eigenmodes containing all waves and instabilities of a force-balanced state is well-organised, and a more general field-theoretical treatment illustrates that the same operators involved fully determine the stability and dynamical response of a plasma at any time in its evolution \citep{demaerel_linear_2016}. In eigenmode computations, perturbations are expanded as normal modes with time dependence $\exp(-i\omega t)$, leading to a generalised eigenvalue problem for the spatial eigenfunctions and complex eigenfrequencies $\omega=\omega_R+i\omega_I$.
The real and imaginary parts carry direct physical meaning: $\omega_R\neq 0$ yields oscillations, while $\omega_I>0$ ($\omega_I<0$) indicates exponential growth (damping), while overstable modes exhibit both oscillation and growth. Spectral theory has been highly successful in the study of laboratory fusion plasmas, and is at the very basis of all helio- or astero-seismological studies.

Non-adiabatic terms added to the (M)HD energy evolution equation, such as radiative losses and (usually) parametrically prescribed heating break the time-reversibility of the ideal system: the neutral entropy mode can move upwards into the unstable half-plane, and introduces a variety of unstable eigenmodes associated with thermal instability. 
In non-uniform magnetised plasmas these include both discrete global modes and a thermal continuum corresponding to singular, extremely localised eigenfunctions \citep{van_der_linden_existence_1991}. 
Such continua are well known in MHD \citep{goedbloed_magnetohydrodynamics_2019} but notably the thermal continuum also arises in purely hydrodynamic, non-adiabatic settings \citep{keppens_hydrodynamic_2025}. 

\citet{van_der_linden_existence_1991} first identified a thermal continuum in a 3D cylindrical coronal loop model, while \citet{van_der_linden_thermal_1991b} showed how finite perpendicular thermal conduction can modify the thermal continuum and can introduce unstable eigenmodes with fine spatial structure, while \citet{ireland_thermal_1992} considered the effects of finite resistivity.
More recently, the \textsc{Legolas} eigenvalue solver \citep{claes_legolas_2020, claes_legolas_2023} has enabled full non-adiabatic spectra to be computed for flowing, magnetised fluxtubes and realistic, stratified atmospheres \citep[e.g.][]{hermans_spectroscopic_2024, keppens_hydrodynamic_2025}.

However, spectral predictions for thermal instability in coronal loops have yet to be confronted with direct time-dependent simulations of the same equilibria.
Key open questions are whether the most unstable eigenmodes are actually excited in initial-value problems, how efficiently continuum modes can be triggered, and how quickly nonlinear evolution causes the system to depart from the linear picture that one may anticipate from the spectral computation.

\subsection{Aims and outline}
We address these questions in a setting that combines the complications of gravitational stratification with thermal instability: a 1D hydrodynamic coronal loop with optically thin radiation and background heating.
Our goal is to perform a unified comparison between three complementary descriptions of the same equilibrium:
\begin{enumerate}
    \item \emph{Spectral analysis} identifying acoustic and thermal modes,
    including the thermal continuum.
    \item \emph{Linear initial-value evolution} using a new
    \textsc{Legolas-IVP} extension to test mode accessibility and measured
    growth rates.
    \item \emph{Nonlinear evolution} with \textsc{MPI-AMRVAC}, determining when
    the system departs from linear behaviour and how condensations form.
\end{enumerate}

To our knowledge this is the first systematic comparison of spectral, linear-IVP, and nonlinear descriptions of thermal instability in a coronal loop context. 
Related recent efforts to unify insights from spectral computations and observed nonlinear evolutions focused on coronal current sheets \citep{Jordi2025} and on eigenmode initialisation of nonlinear simulations for Kelvin-Helmholtz and tearing-unstable configurations \citep{de_jonghe_eigenmode_2026}.
Sect.~\ref{s-methods} describes the governing equations and numerical methods.
Sect.~\ref{s-results} presents the spectral properties, linear responses, and transition to nonlinear evolution.
Sect.~\ref{s-summary} summarises the results.

\section{Model and methods}\label{s-methods}
\subsection{Governing equations}
We model the loop plasma as a fluid governed by hydrodynamics.
Because the solar corona is typically a low plasma beta environment, we adopt the crude but often used approximation in which the magnetic field does not evolve, but the magnetic loop shape dictates all plasma motions. 
This reduces the dimensionality to a field-aligned 1D evolution. 
We will also consider the part of the loop embedded fully in the corona, and hence mimic its coupling to the denser chromosphere at both feet by a perfect-line-tying condition; i.e. take the loop plasma as confined in 1D between two solid-wall boundaries.
This boundary choice enforces vanishing velocity at the domain edges, while still allowing density and temperature variations to evolve self-consistently along the loop.
The fluid is subject to optically thin radiative losses and background heating, while thermal conduction along the loop is deliberately neglected in order to isolate the role of radiative cooling in driving instability. 
By ignoring the intricate thermodynamic coupling with the chromosphere, this study thereby eliminates all TNE processes, and we can concentrate on TI-driven effects in stratified loop settings.
This leads to the following governing equations:
\begin{align}
    &\text{Continuity:} \nonumber \\
    &\frac{\partial \rho}{\partial t} + \frac{\partial}{\partial s}(\rho v) = 0, \label{eqn:cont}\\[0.5em]
    &\text{Momentum:} \nonumber \\
    &\rho\left(\frac{\partial v}{\partial t} + v\frac{\partial v}{\partial s} \right) 
      = - \frac{\partial p}{\partial s} + \rho g_{\parallel}(s), \\[0.5em]
    &\text{Energy:} \nonumber \\
    &\rho \frac{\partial T}{\partial t} = - \rho v \frac{\partial T}{\partial s} 
      - \frac{(\gamma - 1)}{\mathcal{R}}\left[p \frac{\partial v}{\partial s} 
      + \rho \mathcal{L}
      \right]. \label{eqn:energy}
\end{align}
The equations are written with all variables varying along the field-aligned loop coordinate \(s\). 
For simplicity, we have assumed that no variation of the cross-sectional area of the loop along $s$ is at play.
The mass density and velocity are denoted \(\rho\) and \(v\) respectively. 
Gravity is projected onto the loop through $g_\parallel(s)$, as discussed in Sect.~\ref{sec:loop-setup}.
The relation imposed between thermal pressure $p$, and temperature $T$ is the ideal gas law,
\begin{equation}
    p = \mathcal{R} \rho T = \frac{k_{\rm B}} {\mu m_{\rm p}} \rho T,
\end{equation}
where $\mathcal{R}$ denotes the gas constant, with Boltzmann constant $k_{\rm B}$, proton mass $m_{\rm p}$, and mean molecular weight $\mu=1.0$.
We adopt this value for consistency between different numerical codes; it sets the conversion between $\rho$, $n$, and $p$ but does not affect the qualitative instability mechanisms.
The adiabatic index is taken to be $\gamma = 5/3$, appropriate for a monatomic gas. 

The net heat-loss term \(\mathcal{L}\) is the difference between energy losses due to optically thin radiative emission, and gains due to a prescribed heating term. 
It appears as a sink or source term in the energy equation (\ref{eqn:energy}) as 
\begin{equation} 
\rho \mathcal{L}(\rho,T)=n^2\Lambda(T)-H(s), 
\end{equation} 
where $n=\rho/(\mu m_{\rm p})$ is the number density, and the background heating rate $H(s)$ is spatially-varying and constant in time. 
We adopt the convention of \citet{field_thermal_1965}, such that \(\mathcal{L}>0\) corresponds to net cooling  -- note that this is the opposite of \citet{keppens_hydrodynamic_2025}. 
Note that the volumetric terms \(n^2\Lambda(T)\) and \(H(s)\) both have units of erg~cm\(^{-3}\)~s\(^{-1}\), while \(\mathcal{L}\) is defined per unit mass, with units of erg~g\(^{-1}\)~s\(^{-1}\).
The cooling curve $\Lambda(T)$ employed is a composite, implemented in both \textsc{Legolas} and \textsc{MPI-AMRVAC}, consisting of the tabulation by \citet{colgan_radiative_2008} extended to low temperatures by merging the curve of \citet{dalgarno_heating_1972}. 
We note that both codes actually share the many options available for the cooling table, and how this choice affects both linear and nonlinear evolutions in simple periodic, initially homogeneous settings was already explored in \cite{hermans_effect_2021}.

Even when we simulate the full nonlinear set of equations, we can split all quantities into a stationary background state and a perturbed part, indicated with a subscript 0 and 1 respectively; e.g.
\begin{equation}
    \rho(s,t) = \rho_0(s) + \rho_1(s,t),
\end{equation}
and similarly for $p$ and $T$.
The background is adopted as stationary (i.e. \(v_0=0\)), in order to isolate the role of thermal instability in an otherwise hydrostatic loop.
The equilibrium is furthermore taken to be isothermal, yielding a stratified hydrostatic atmosphere, and is thermally balanced by enforcing \(\mathcal{L}_0=\mathcal{L}(\rho_0,T_0)=0\) everywhere.
Enforcing thermal balance, with heating offsetting radiative losses, ensures that any evolution is triggered purely by perturbations rather than by thermodynamically driven drift of the equilibrium state.
This is also the natural prescription for the linear framework, as both \textsc{Legolas} and \textsc{Legolas-IVP} treat all background quantities, including $\mathcal{L}_0$, as fixed and do not account for $d\mathcal{L}_0/dt$. 
If $\mathcal{L}_0 \neq 0$, the background would evolve on its own slow thermal timescale as the net heating or cooling drives changes in the background, causing $\mathcal{L}_0$ to become time-dependent. 
By enforcing $\mathcal{L}_0 = 0$ everywhere, this drift is eliminated exactly.
This implies a spatially varying heating profile, since \(H_0(s)=n_0^2(s)\Lambda(T_0)\) for the adopted isothermal background.
Thus, the equilibrium heating is height-dependent in our model, but this dependence is not imposed as an additional ad hoc heating prescription; rather, it is fully determined by the requirement of local thermal balance in the chosen background state. 
This differs from the common practice in multidimensional MHD models of prescribing an independent empirical height-dependent heating function (e.g. footpoint-concentrated heating) \citep[e.g.][]{Brughmans2022,Fang2015}.

\subsection{Loop setup}\label{sec:loop-setup}
We consider a semi-circular coronal loop of length \(L = 50\)~Mm.
The background temperature is fixed at a typical coronal value of \(T_0 = 10^6\)~K everywhere along the loop.
Gravity is taken as the solar surface value \(g_\odot = 274~\mathrm{m\,s^{-2}}\) and is projected onto the field-aligned coordinate \(s\) via
\begin{equation}
    g_\parallel(s) = -\,g_\odot \cos\!\left(\frac{\pi s}{L}\right),
\end{equation}
where \(s=0\) and \(s=L\) correspond to the footpoints and \(s=L/2\) to the apex.
The equilibrium density \(\rho_0(s)\) is obtained from the isothermal hydrostatic balance condition
\begin{equation}
    \frac{dp_0}{ds}
    = c_i^2\,\frac{d\rho_0}{ds}
    = \rho_0\, g_\parallel(s),
\end{equation}
with the isothermal sound speed \(c_i^2 = k_{\rm B}T_0/(\mu m_{\rm p})\).
This yields
\begin{equation}
    \rho_0(s)
    = \rho_{\mathrm{fp}}\,
      \exp\!\left[
        -\,\frac{g_\odot L}{\pi c_i^2}
        \,\sin\!\left(\frac{\pi s}{L}\right)
      \right].
\end{equation}
Here \(\rho_{\mathrm{fp}} = 2 \times 10^{-15}~\mathrm{g\,cm^{-3}}\) is the reference footpoint density.
All quantities are non-dimensionalised using the reference scales \(\bar{L} = 10^9\)~cm, \(\bar{n} = 10^9~\mathrm{cm^{-3}}\), and \(\bar{T} = 10^6\)~K.
The corresponding time unit is \(\bar{t} = \bar{L} / c_i \approx 110~\mathrm{s}\).

\subsection{Numerical methods}
\subsubsection{\textsc{Legolas}: spectral solver}
To compute the linear stability properties of the loop equilibrium we use the \textsc{Legolas} eigenvalue code \citep{claes_legolas_2020, claes_legolas_2023}, which solves the full non-adiabatic (M)HD eigenvalue problem for equilibria varying along a single spatial coordinate. 
Here we adopt the one-dimensional hydrodynamic limit, where the perturbed state vector is $(\rho_1, v_1, T_1)$.

The linearised equations are discretised using a finite-element method on a uniform grid of \(N=200\) elements with quadratic and cubic basis functions. 
This yields a sparse generalised eigenvalue problem
\begin{equation}
    \mathbf{A} \mathbf{x} = \omega\, \mathbf{B} \mathbf{x}, \label{q-eig}
\end{equation}
whose solutions give the complex eigenfrequencies $\omega$ and associated spatial eigenfunctions. 
Fixed-wall (line-tied) boundary conditions are imposed at both footpoints to model a coronal loop section that is tied to the dense chromosphere, enforcing $v_1=0$ at the boundaries while allowing $\rho_1$ and $T_1$ to evolve self-consistently. Since there is no ignorable coordinate, there is no wavevector to prescribe, and this spectral analysis will yield all combinations of eigenfrequencies-eigenvectors $[\omega; \ \rho_1^{\omega}(s), v_1^{\omega}(s),  T_1^{\omega}(s)]$ that are physical eigenmodes of the loop setup. All these quantities are generally complex-valued, giving the natural response for a $\exp(-i\omega t)$ normal mode variation.

\subsubsection{\textsc{Legolas-IVP}: linear time evolution}\label{sec:legolas-ivp}
To complement the spectral analysis, we use a newly developed initial-value solver module for \textsc{Legolas}, referred to as \textsc{Legolas-IVP}, which integrates the same linearised equations directly in time.
The IVP solver uses the same finite-element spatial discretisation as the eigenvalue problem, allowing spectral and time-dependent results to be compared directly without changing the underlying numerical representation of the equilibrium.

We evolve the linearised system on the same uniform mesh of \(N=200\) finite elements as in the spectral calculations, and advance in time using a second-order implicit midpoint scheme.
Unless stated otherwise, all runs start from the thermally balanced equilibrium \(\mathcal{L}(\rho_0,T_0)=0\) with \(v_1(s,0)=0\), and are evolved for \(9\,\bar t\), approximately corresponding to \(t_{\rm end}=1000~\mathrm{s}\).
This duration covers several exponential growth times of the unstable thermal branch while remaining within the linear regime (see Sect. \ref{sec:growth_rates}).
Note that \textsc{Legolas-IVP} works in real space (not in the Fourier space used for the spectral analysis), and hence needs both boundary and initial values for all (real-valued) variables $[\rho_1(s,t),v_1(s,t),T_1(s,t)]$.

Initial conditions are prescribed as small-amplitude, localised Gaussian pulses applied to \(\rho_1\) and \(T_1\).
Each pulse has a central peak position \(s_0\) and has standard deviation \(\sigma\),
\begin{equation}
    \delta(s) = A \exp\!\left[-\left(\frac{s-s_0}{\sigma}\right)^2\right], \qquad |A|\ll 1,
\end{equation}
where \(A=0.01\) is chosen such that relative perturbations remain in the linear regime.
Unless stated otherwise, we use a reference width \(\sigma=3~\mathrm{Mm}\); for the mode-selectivity experiments in Sect.~\ref{sec:growth_rates} we additionally employ a narrower pulse (\(\sigma=0.5~\mathrm{Mm}\)) to preferentially excite individual localised continuum eigenfunctions.

We consider three thermodynamic classes of perturbations.
For \textit{isochoric} pulses we set \(\rho_1(s,0)=0\) and \(T_1(s,0)=T_0(s)\,\delta(s)\).
For \textit{isobaric} pulses we impose \(\rho_1(s,0)=-\rho_0(s)\,\delta(s)\) and \(T_1(s,0)=T_0(s)\,\delta(s)\), giving \(p_1(s,0)=0\).
For \textit{isentropic} pulses we choose coupled perturbations such that \(S_1(s,0)=0\), where we employ the entropy variable $S=p\rho^{-\gamma}$.
By imposing these controlled perturbations, the time-dependent simulations enable us to assess whether this polarisation of the initial perturbation preferentially selects and initiates certain normal modes and to extract growth rates directly from the temporal evolution for comparison with the \textsc{Legolas} spectrum. 
The implementation of this linear initial-value solver, including verification and grid-convergence tests, is documented in Appendix~\ref{app:ivp-verification}.

\subsubsection{\textsc{MPI-AMRVAC}: nonlinear hydrodynamics}
Nonlinear evolution is studied using \textsc{MPI-AMRVAC} \citep{xia_mpi-amrvac_2018, keppens_mpi-amrvac_2021,keppens_mpi-amrvac_2023}, a high-resolution shock-capturing finite-volume code for (magneto)hydrodynamics.
Although the code supports adaptive mesh refinement (AMR), in this work we employ uniform 1D grids. 
For the growth-rate comparison runs (Sect.~\ref{sec:growth_rates}) we use $N=16384$ cells, giving a physical resolution of $\Delta s \simeq 3.1$~km, while the full nonlinear condensation run (Sect.~\ref{sec:nonlinear_transition}) uses $N=65536$ cells ($\Delta s \simeq 0.76$~km) to resolve the steep density and temperature gradients generated by thermal runaway and condensation.

\textsc{MPI-AMRVAC} solves the same 1D hydrodynamic equations as in \textsc{Legolas}, but in conservative form with state vector $(\rho, m, e)$, where $m=\rho v$ is the field-aligned momentum density and $e$ is the total energy density.
Optically thin radiation, background heating, and field-aligned gravity are included through source terms.
The equilibrium is implemented in split form \citep{yadav_3d_2022}, writing each conserved quantity as a fixed background plus a time-dependent perturbation and evolving $(\rho_1, m_1, e_1)$ about the hydrostatic equilibrium.
This facilitates direct comparison with the linear (\textsc{Legolas}) calculations performed about the same state.

Spatial fluxes are computed using the Harten--Lax--van Leer (HLL) approximate Riemann solver \citep{harten_upstream_1983} with a second-order van Leer limiter \citep{van_leer_towards_1974}, and time integration is performed with a three-stage strong-stability-preserving Runge--Kutta scheme (SSP-RK3) \citep{gottlieb_strong_2001}.
The same radiative loss function and heating profile are implemented as in the linear calculations, enabling like-for-like comparisons of growth rates and mode development.

At both footpoints we impose reflecting (solid-wall) boundary conditions via a ghost-cell prescription.
We apply symmetric boundary conditions for the perturbations $(\rho_1, e_1)$ and antisymmetric conditions for the momentum perturbation $m_1$.
In practice, the symmetric prescription enforces a zero normal gradient at the boundary, while antisymmetry enforces a sign change across the boundary and implies $m_1=0$ at $s=0$ and $s=L$, corresponding to zero mass flux and vanishing velocity at the walls.

\section{Results and discussion}\label{s-results}
\subsection{The eigenmode spectrum of a stratified loop}

\begin{figure*}[htbp]
    \centering
    \includegraphics[width=\linewidth]{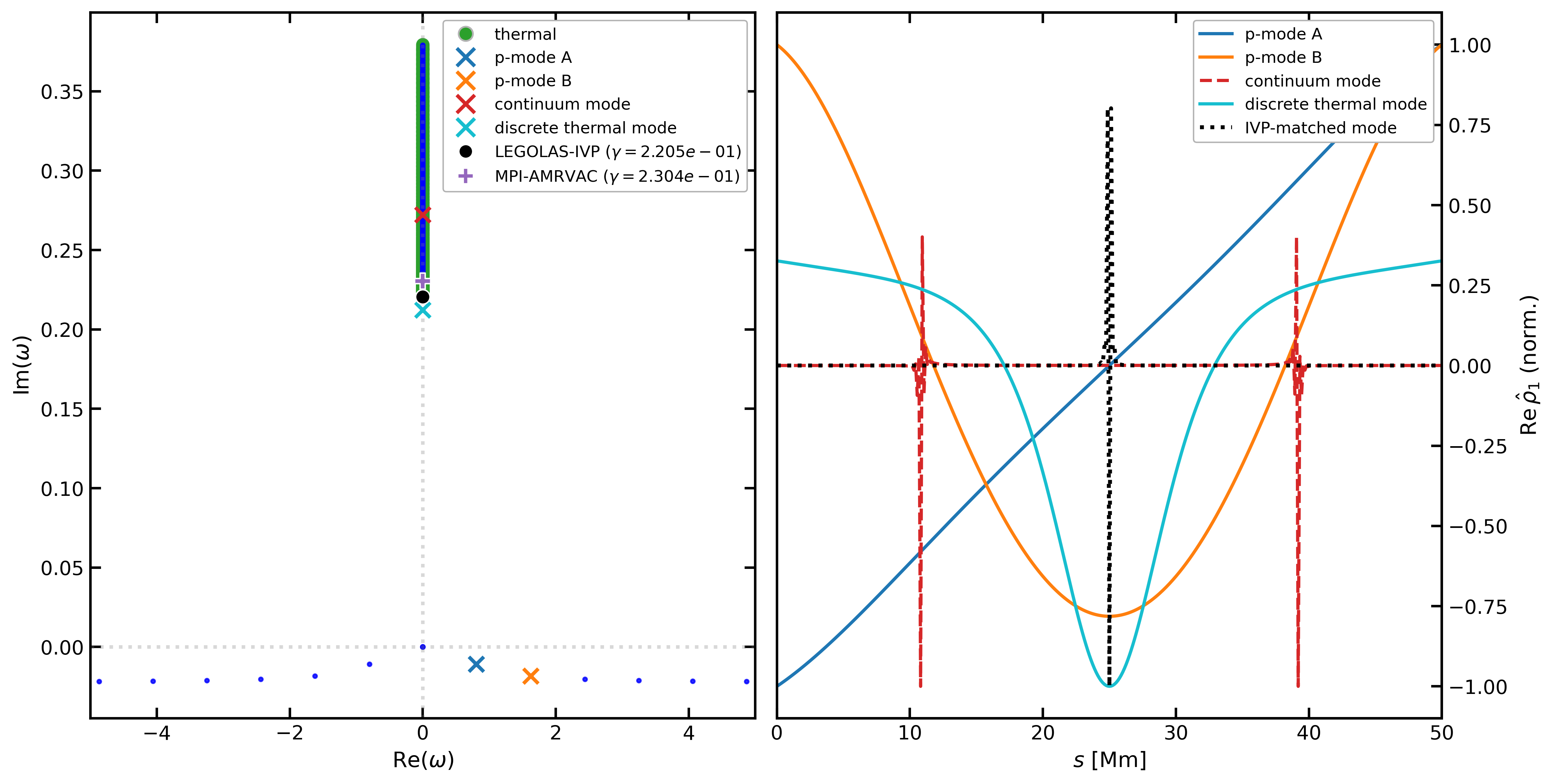}
    \caption{Linear spectrum and selected eigenfunctions for the thermally balanced 1D loop.
    \textit{Left:} Eigenvalue spectrum obtained with \textsc{Legolas}, showing acoustic modes, the thermal continuum, and an isolated discrete thermal mode. 
    Blue dots denote individual eigenmodes. Selected modes are highlighted by coloured crosses: two acoustic p-modes (blue, orange), a thermal-continuum mode (red), and a discrete thermal mode (cyan). 
    The measured growth rates obtained from \textsc{Legolas-IVP} and \textsc{MPI-AMRVAC} for the isobaric perturbation (cf. Fig.~\ref{fig:growth-rates}) are overplotted as black and purple markers, respectively, on the $\Re(\omega)=0$ axis.
    \textit{Right:} Real parts of the corresponding density eigenfunctions along the loop coordinate \(s\), normalised to unit amplitude. 
    The coloured curves correspond to the four highlighted eigenmodes in the left panel. 
    The black dotted curve shows the eigenfunction of the \textsc{Legolas} mode whose eigenfrequency corresponds to the growth rate measured in the \textsc{Legolas-IVP} simulation.}
    \label{fig:loop-spectrum}
\end{figure*}

Fig.~\ref{fig:loop-spectrum} (left) shows the eigenfrequency spectrum computed by \textsc{Legolas} for the thermally balanced loop equilibrium ($\mathcal{L}_0=0$), together with representative real parts of the density eigenfunctions (right).
The spectrum separates into three main families of solutions: two branches of acoustic (p-) modes, and a thermal branch, in agreement with the analytical results of \citet{keppens_hydrodynamic_2025}. 

The vertical band of modes along $\Re(\omega)=0$ corresponds to the thermal continuum.
These modes have purely imaginary frequencies spanning a continuous range corresponding to $M=0$, where
\begin{equation}
 M \equiv \omega c_s^{2} + i(\gamma - 1)\left(\mathcal{L}_0 + \rho_0 \mathcal{L}_{\rho} - T_0 \mathcal{L}_{T}\right).  \label{q-mfac}
\end{equation}
Here, \(c_s\) denotes the (local) adiabatic sound speed, and \(\mathcal{L}_{\rho}\) and \(\mathcal{L}_{T}\) are the partial derivatives of the heat-loss function with respect to density and temperature, respectively, i.e.
\(\mathcal{L}_{\rho}\equiv (\partial \mathcal{L}/\partial \rho)_T\) and
\(\mathcal{L}_{T}\equiv (\partial \mathcal{L}/\partial T)_\rho\), evaluated on the equilibrium state \((\rho_0,T_0)\).

The continuum eigenfunctions are strongly localised in space, reflecting the singular behaviour expected of continuum modes in non-uniform, non-adiabatic systems.
Continuum eigenfunctions, one of which is illustrated by the red curve in the right panel, will always appear as extremely localised rather than truly singular, due to the finite nature of the numerical grid.
Physically, the continuum modes are a generalisation of the classical thermal instability to stratified media.

Pairs of modes symmetric about the imaginary axis correspond to standing acoustic oscillations.
These have non-zero real parts and follow the expected p-mode ordering, with frequencies increasing with the number of nodes in the eigenfunction, which could be associated with a loop-aligned corresponding wavenumber (even though no actual simple Fourier representation is possible due to the stratification).
Their eigenfunctions display the expected nodal structure of a Sturmian sequence, reflecting the fact that in the adiabatic limit, the governing equation can be recast into Sturm--Liouville form \citep{keppens_hydrodynamic_2025}.
Radiative losses introduce a weak damping, shifting the eigenfrequencies slightly into the lower half-plane without altering their oscillatory character.
The blue and orange curves in the right panel show examples of these p-mode eigenfunctions.

Finally, we find a single discrete thermal mode with a purely imaginary eigenfrequency (cyan marker).
This global mode lies just below the continuum and is the most unstable non-singular mode.
Its spatial structure is extended across the loop, with the largest amplitudes near the apex where the background density is lowest. 
That such an isolated global mode is possible is related to the fact that the thermal continuum is displaying a local extremum at the apex of the loop. 
Whether such discrete modes exist for a particular loop equilibrium can be analysed using WKB techniques, demonstrated in the more complex MHD case with flow in \citet{hermans_spectroscopic_2024}. 

Together, these branches define the complete linear response of the system and provide the reference against which we compare the linear initial-value simulations (\textsc{Legolas-IVP}) and the fully nonlinear hydrodynamic simulations (\textsc{MPI-AMRVAC}).
As shown in the next section, the time-dependent simulations recover exactly the same dynamical components: no additional linear behaviour emerges beyond what is already contained in the spectrum.

\subsection{Time-dependent linear evolutions in a 1D coronal loop}\label{sec:time-dep-sim}

\begin{figure*}[htb]
    \centering
    \includegraphics[width=1.0\linewidth]{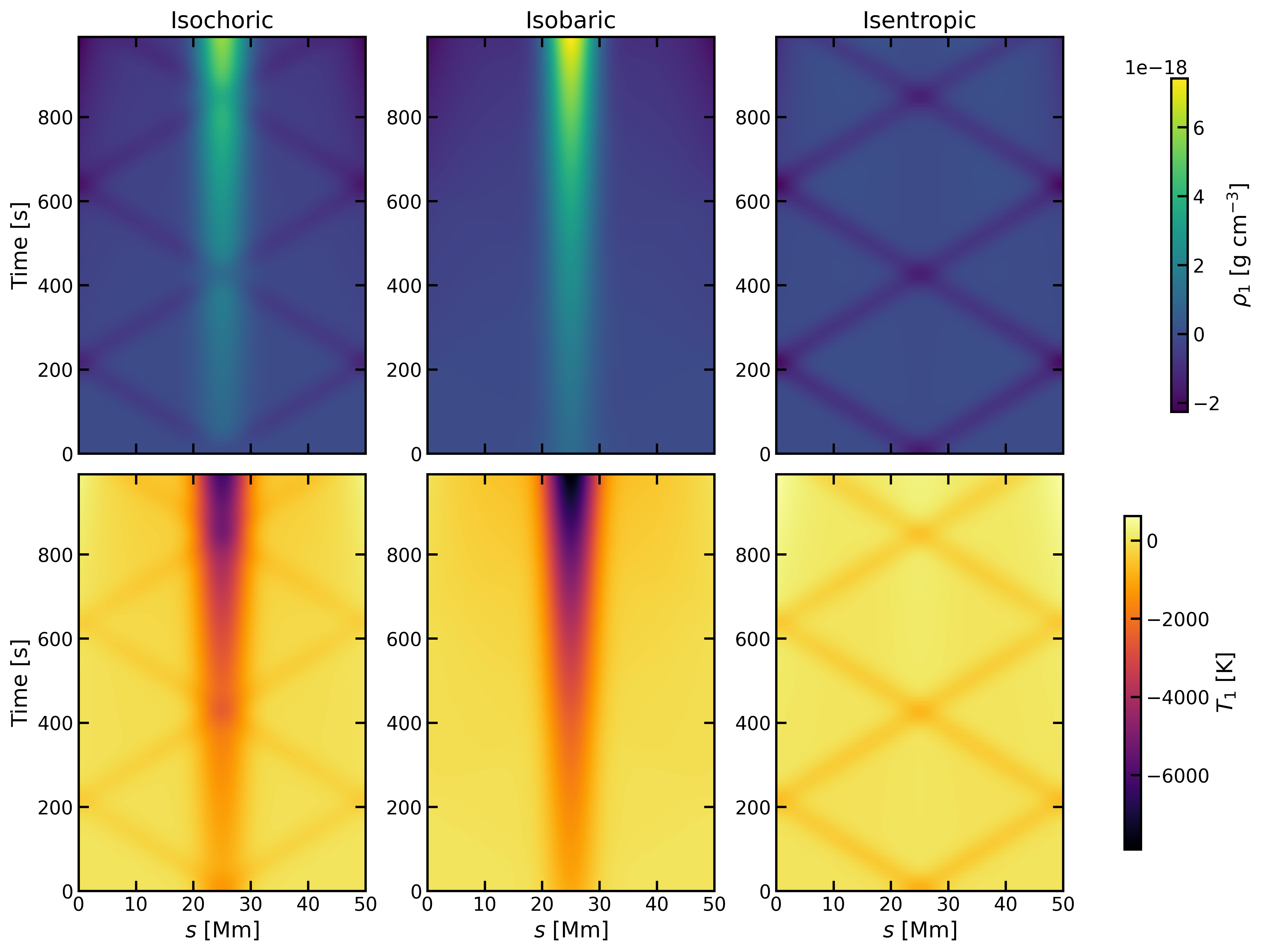}
    \caption{Space--time maps of density (top row) and temperature (bottom row) perturbations for the three initial pulses: isochoric (left), isobaric (middle), and isentropic (right). Each panel shows the perturbation amplitude as a function of position along the loop and simulation time. Note how a centrally cooling, denser region at the apex develops for both isochoric and isobaric pulses, corresponding to the thermal instability, which is not triggered by an isentropic pulse.}
    \label{fig:pulses-evolution}
\end{figure*}

\begin{figure*}[htb]
    \centering
    \includegraphics[width=1.0\linewidth]{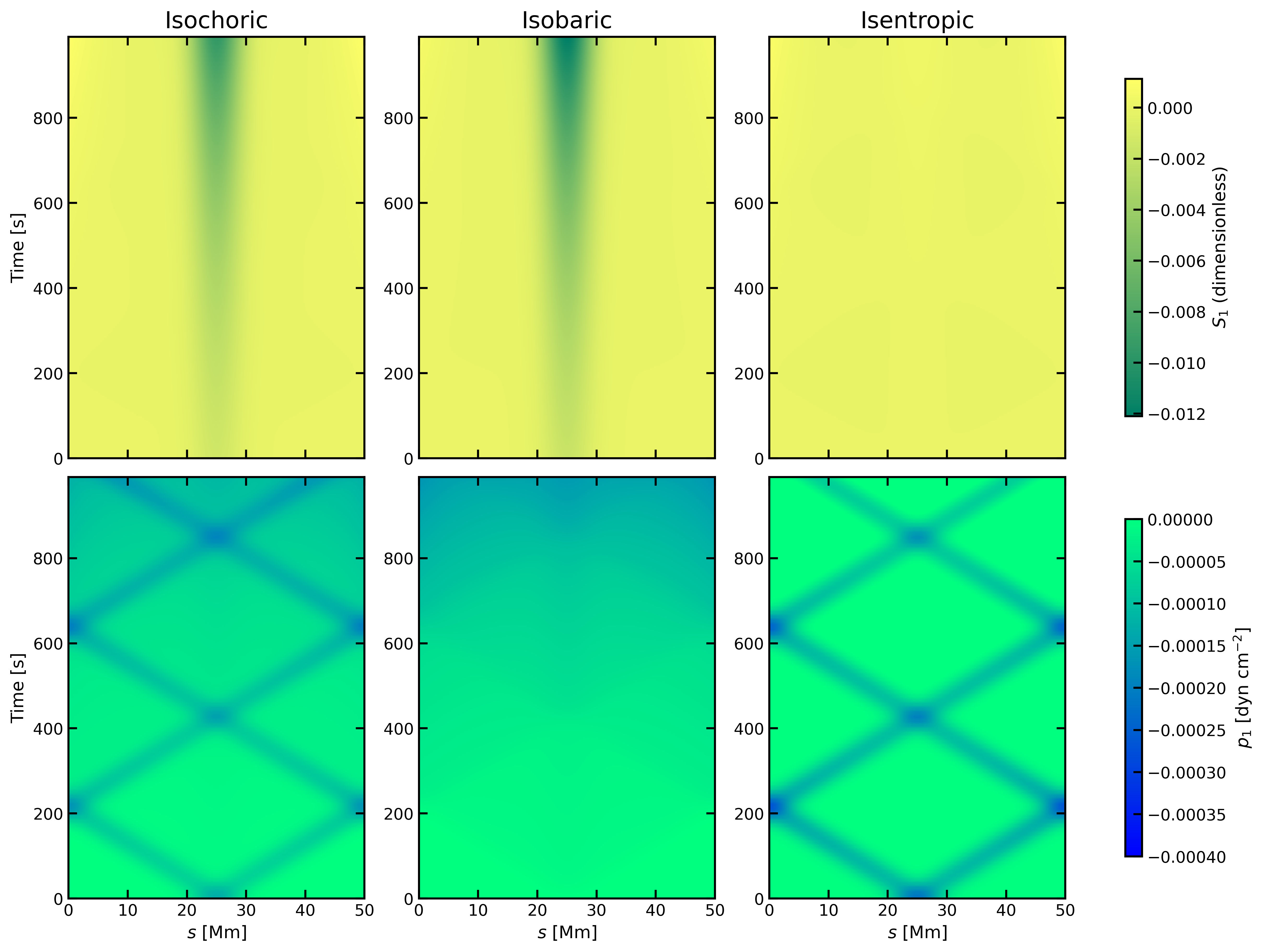}
    \caption{Space--time evolution of the entropy perturbation $S_1$ (top row) and pressure perturbation $p_1$ (bottom row) for isochoric (left), isobaric (middle), and isentropic (right) initial pulses. Colour scales indicate the perturbation magnitude in each case. As expected, the isentropic initial pulse maintains the original equilibrium entropy variation.}
    \label{fig:pulses-derived}
\end{figure*}

It is not obvious a priori which parts of the spectrum can actually be excited in time-dependent simulations. 
Controlled perturbations therefore provide a useful test case: they allow us to identify which spectral branches manifest in practice, which thermodynamic constraints favour their excitation, and to what extent the expected mode structure is recovered. 
To this end we compare the spectral predictions with linear initial-value simulations in \textsc{Legolas-IVP}. 
The corresponding nonlinear \textsc{MPI-AMRVAC} runs will later be used to confirm growth rates and follow the departure from linear behaviour, but the controlled IVP tests already clarify the physical relevance of specific spectral branches.

Fig.~\ref{fig:pulses-evolution} shows the space-time evolution of density and temperature perturbations for the three representative thermodynamic pulses introduced in Sect.~\ref{sec:legolas-ivp}: isochoric, isobaric, and isentropic.
The isochoric pulse triggers local thermal instability by exciting the thermal continuum modes, as well as triggering acoustic propagating waves, since a pure temperature perturbation necessarily induces a pressure disturbance and launches p-modes.
The isobaric pulse, constructed to satisfy $p_1 = 0$, suppresses acoustic oscillations entirely and isolates the exponentially-growing local thermal instability.
The isentropic pulse excites only weakly damped acoustic oscillations, and shows no sign of thermal runaway. 
When acoustic modes are excited, we see them propagate to the loop footpoints, from where they get reflected as a result of the solid-wall boundary assumption there. This process repeats, while their amplitude dampens weakly, due to the non-adiabatic effects.

Fig.~\ref{fig:pulses-derived} shows the evolution of the pressure and entropy perturbations for the very same initial and boundary value experiments.
Isochoric and isobaric constraints are imposed only at $t=0$ and are not necessarily preserved in the subsequent evolution.
Nevertheless, in the isobaric case $p_1$ remains small and the response stays close to isobaric over the times shown.
The isentropic pulse remains nearly adiabatic, while the thermally unstable cases exhibit a clear decrease in entropy, reflecting the non-adiabatic nature of the thermal mode.

\begin{figure}[htb]
    \centering
    \includegraphics[width=1\linewidth]{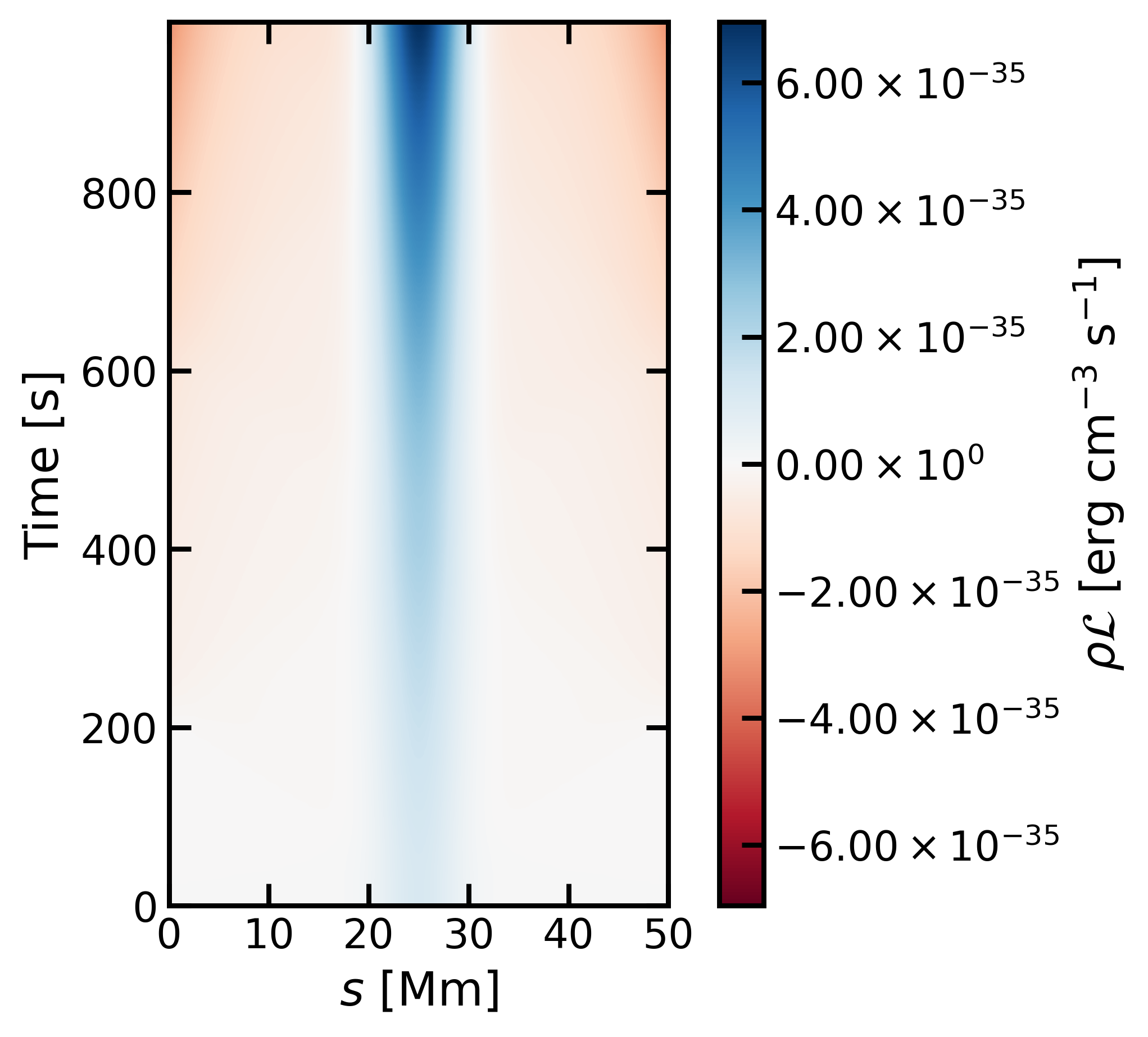}
    \caption{Space--time map of the net heat-loss function $\rho(s)\mathcal{L}(\rho,T) = \rho^{2}(s)\Lambda[T(s)] - H(s)$ for the isobaric initial pulse. The colour map denotes the local value of $\rho\mathcal{L}$ over the loop as a function of time, with blue corresponding to net cooling, and red to net heating.}
    \label{fig:L_heatmap}
\end{figure}

\begin{figure}[h!]
    \centering
    \includegraphics[width=1\linewidth]{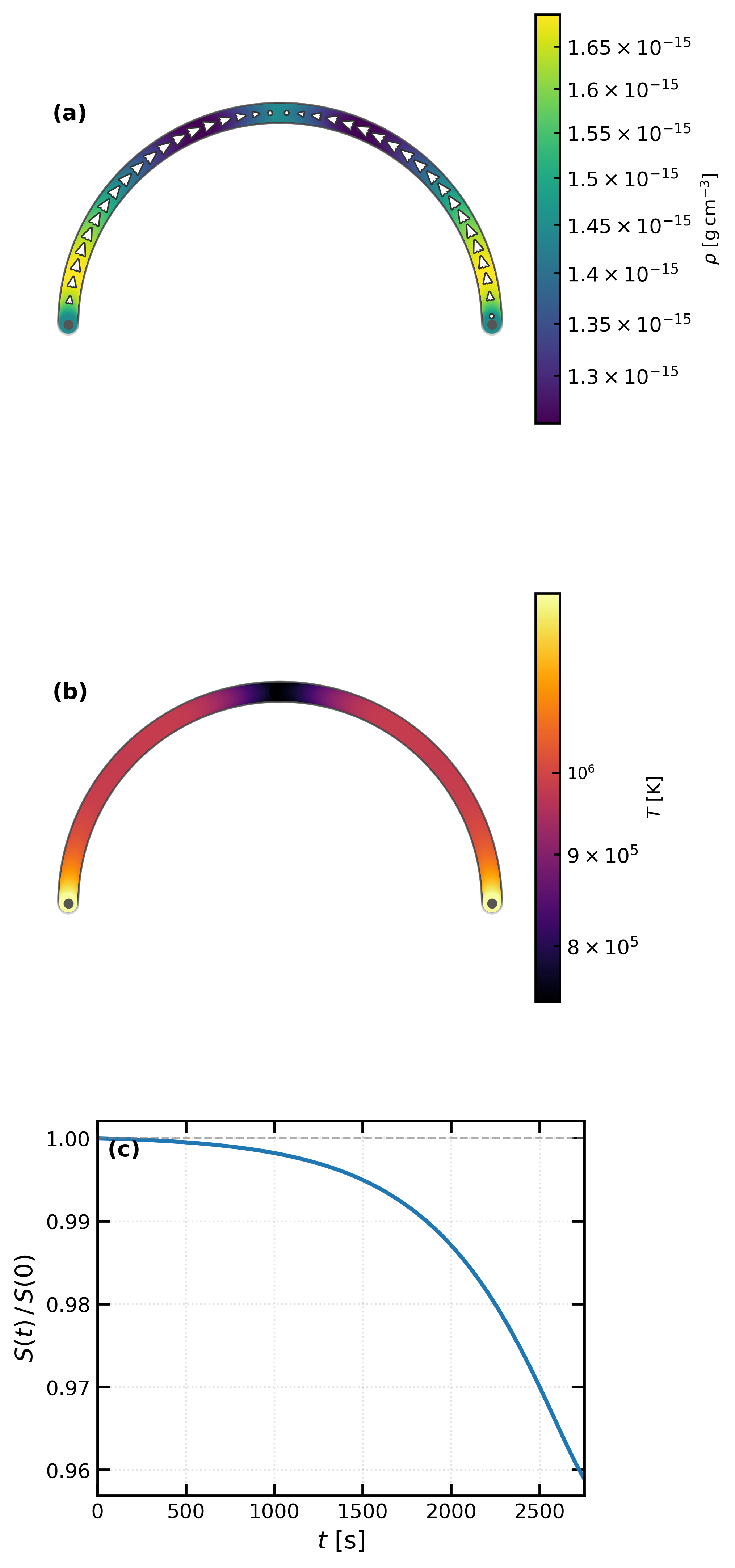}
    \caption{Total density $\rho$ (top) and temperature $T$ (middle) along the loop at $t \approx 2750$~s in the extended isobaric simulation, shown in physical units. 
    The velocity $v$ is overlaid as white arrows in panel~(a), illustrating the converging siphon-like flows directed toward the apex condensation. 
    Panel~(c) shows the loop-integrated total entropy $S(t) = \int (p/\rho^{\gamma})\,\mathrm{d}s$, normalised to its initial value, as a function of time. 
    The entropy decreases monotonically throughout the linear evolution, consistent with the non-adiabatic nature of the dominant thermal continuum mode. }
    \label{fig:anim-end}
\end{figure}

Fig.~\ref{fig:L_heatmap} shows the evolution of the volumetric net heat-loss term \(\rho\mathcal{L}(\rho,T)=\rho^{2}\Lambda(T)-H(s)\) for the isobaric pulse, reconstructed from the linear \textsc{Legolas-IVP} solution to interpret the local thermal imbalance. 
Note that \(\rho\mathcal{L}\) is evaluated from the total fields \(\rho=\rho_0+\rho_1\) and \(T=T_0+T_1\), whereas the \textsc{Legolas} solver is only aware of the linearised source term \((\rho\mathcal{L})_1\).
The loop initially satisfies thermal equilibrium, so $\rho\mathcal{L}=0$ everywhere at $t=0$. 
When the perturbation is applied, the local balance is broken and $\rho\mathcal{L}$ becomes positive (blue) around the apex, indicating excess radiative cooling relative to the imposed heating. 
As the instability grows, the response remains approximately isobaric (cf. Fig.~\ref{fig:pulses-derived}): the temperature decreases while the density increases, driving converging siphon-like flows from the line-tied footpoints towards the cooling site (illustrated by arrows denoting velocity in Fig.~\ref{fig:anim-end}), leaving underdense regions around the footpoints. 
Because optically thin radiative losses scale as $\propto \rho^2 \Lambda(T)$, the mass-depleted regions radiate less efficiently and tip into a state of net heating (red), since $H(s)$ is time-independent. 

Fig.~\ref{fig:anim-end} shows the total density and temperature along the loop at $t \approx 2750$~s, together with the loop-integrated total entropy as a function of time, for the extended isobaric simulation. 
The total entropy $S(t) = \int (p/\rho^{\gamma})\,\mathrm{d}s$, normalised to its initial value, decreases monotonically throughout the linear evolution, confirming that the dominant response is a non-adiabatic thermal mode.
The simulation is terminated at $t \approx 2750$~s, before the perturbation amplitudes start to become comparable to the background.

These controlled tests demonstrate that
(i) the localised thermally unstable continuum eigenmodes are readily excited by suitable density-temperature perturbations;
(ii) purely acoustic (isentropic) perturbations do not trigger thermal instability in a balanced loop; and
(iii) no additional linear behaviour appears beyond the branches identified in the spectrum.
This establishes a one-to-one correspondence between the spectral structure and the time-dependent linear response of the loop.

Even in this minimal gravitationally-stratified setup, siphon-like upflows emerge naturally as a linear response to thermal imbalance.
That these features already appear in the earliest stages of the instability suggests that the linear continuum mode dynamics play a crucial role in explaining prominence formation.
Recently the role of these TI-induced siphon flows was emphasised in the full 3D MHD simulation of a forming quiescent prominence by \citet{donne_mass_2024}.

\subsection{Comparing growth rates and mode structures}
\label{sec:growth_rates}
\begin{figure}[htb]
    \centering
    \includegraphics[width=1\linewidth]{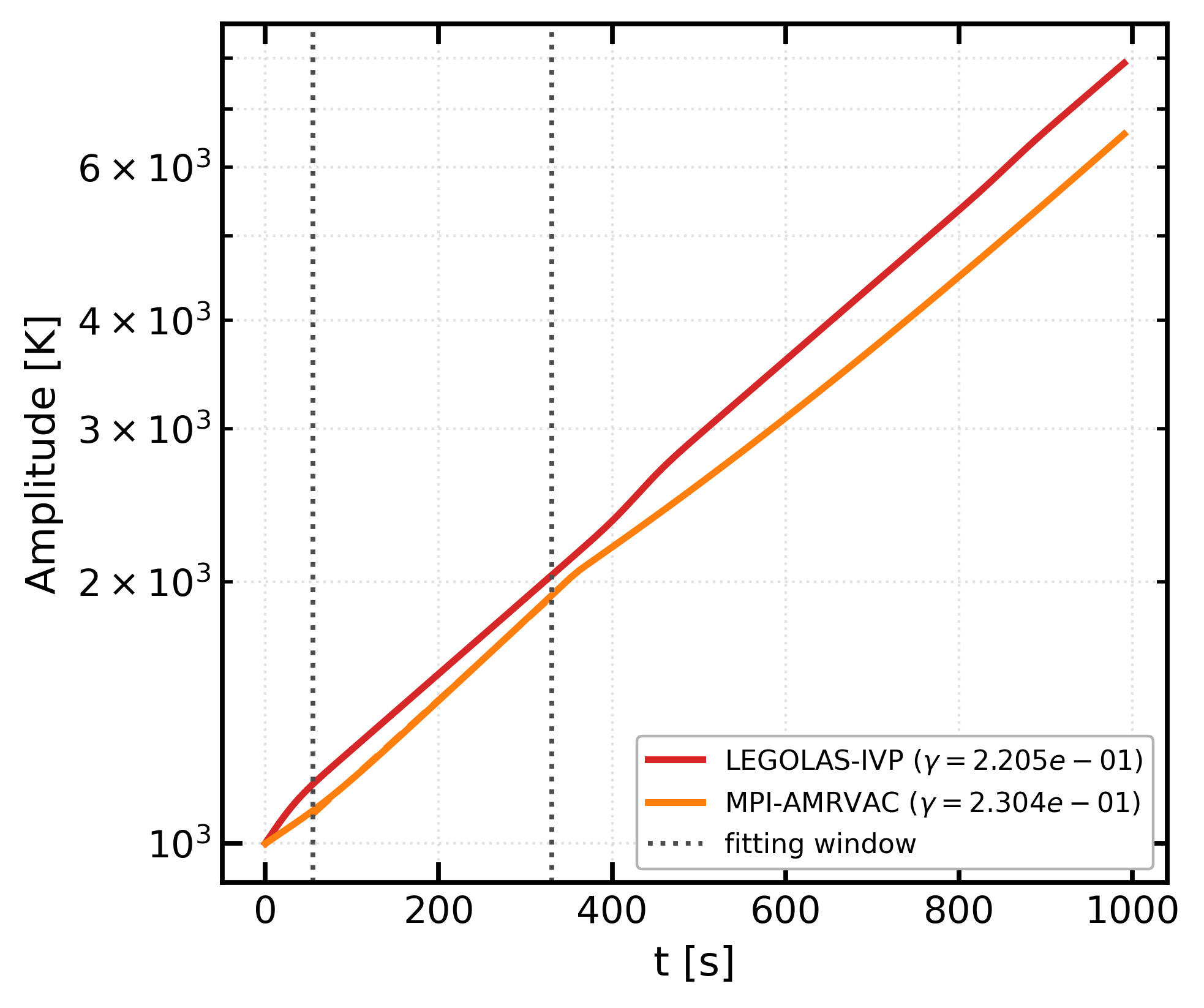}
    \caption{Comparison of spectral and measured thermal-instability growth rates for an isobaric perturbation.
    Evolution of $A(t)=\max_{s\in\mathrm{ROI}}|T_1(s,t)|$ for \textsc{Legolas-IVP} and \textsc{MPI-AMRVAC}. Vertical dotted lines mark the fitting interval.}
    \label{fig:growth-rates}
\end{figure}

We now quantify the instability development by extracting exponential growth rates from time-dependent simulations and comparing them with the thermally unstable branch of the \textsc{Legolas} spectrum.
Since the isobaric pulse suppresses the acoustic response most effectively, we focus on this case as the cleanest representation of the thermal modes. Besides the time-dependent \textsc{Legolas-IVP} evolution, we now also run \textsc{MPI-AMRVAC} with exactly the same initial and boundary conditions.

As an amplitude measure we use the maximum absolute temperature perturbation in a region of interest (ROI) around the initial pulse centre,
\begin{equation}
    A(t) \equiv \max_{s\in \mathrm{ROI}} |T_1(s,t)|,
\end{equation}
where we take $\mathrm{ROI}=[s_0-w,\,s_0+w]$ with half-width $w=6$~Mm and pulse centre $s_0$.
For our reference Gaussian pulse with standard deviation $\sigma=3$~Mm this corresponds to $w=2\sigma$, which brackets the perturbed apex region and prevents the diagnostic from being contaminated by acoustic features outside the condensation site.

During the linear phase we expect $A(t)\propto \exp(\gamma t)$, and we estimate $\gamma$ from a linear regression of $\ln A(t)$ over the interval $t\in[55,330]$~s (indicated in Fig.~\ref{fig:growth-rates}).
We report $\gamma$ in inverse code-time units (with $\bar t=\bar L/c_i$), so that the corresponding physical e-folding time is $\tau_e=\bar t/\gamma$.
The e-folding time $\tau_e$ is the time required for the perturbation amplitude to increase by a factor of $e$, and therefore provides a direct timescale for instability evolution.

Figure~\ref{fig:growth-rates} shows that both \textsc{Legolas-IVP} and \textsc{MPI-AMRVAC} exhibit a clear exponential growth during the early phase and yield growth rates consistent with the unstable thermal branch of the \textsc{Legolas} spectrum.
The fitted values are $\gamma_{\rm IVP}=2.205\times10^{-1}$ and $\gamma_{\rm AMRVAC}=2.304\times10^{-1}$, corresponding to $\tau_e\simeq 499$~s and $477$~s respectively.
Because the thermal continuum contains a continuous range of unstable eigenvalues with (numerically) extremely localised eigenfunctions, a finite-width Gaussian pulse necessarily excites a band of neighbouring continuum eigenfunctions rather than a single eigenfunction.
We therefore compare the fitted \(\gamma\) with the thermal-continuum eigenvalue whose eigenfunction is localised closest to the pulse centre \(s_0\); for the apex-centred pulse in Fig.~\ref{fig:growth-rates} this corresponds to the local continuum growth rate near the loop apex.

\begin{figure}[htb]
    \centering
    \includegraphics[width=1\linewidth]{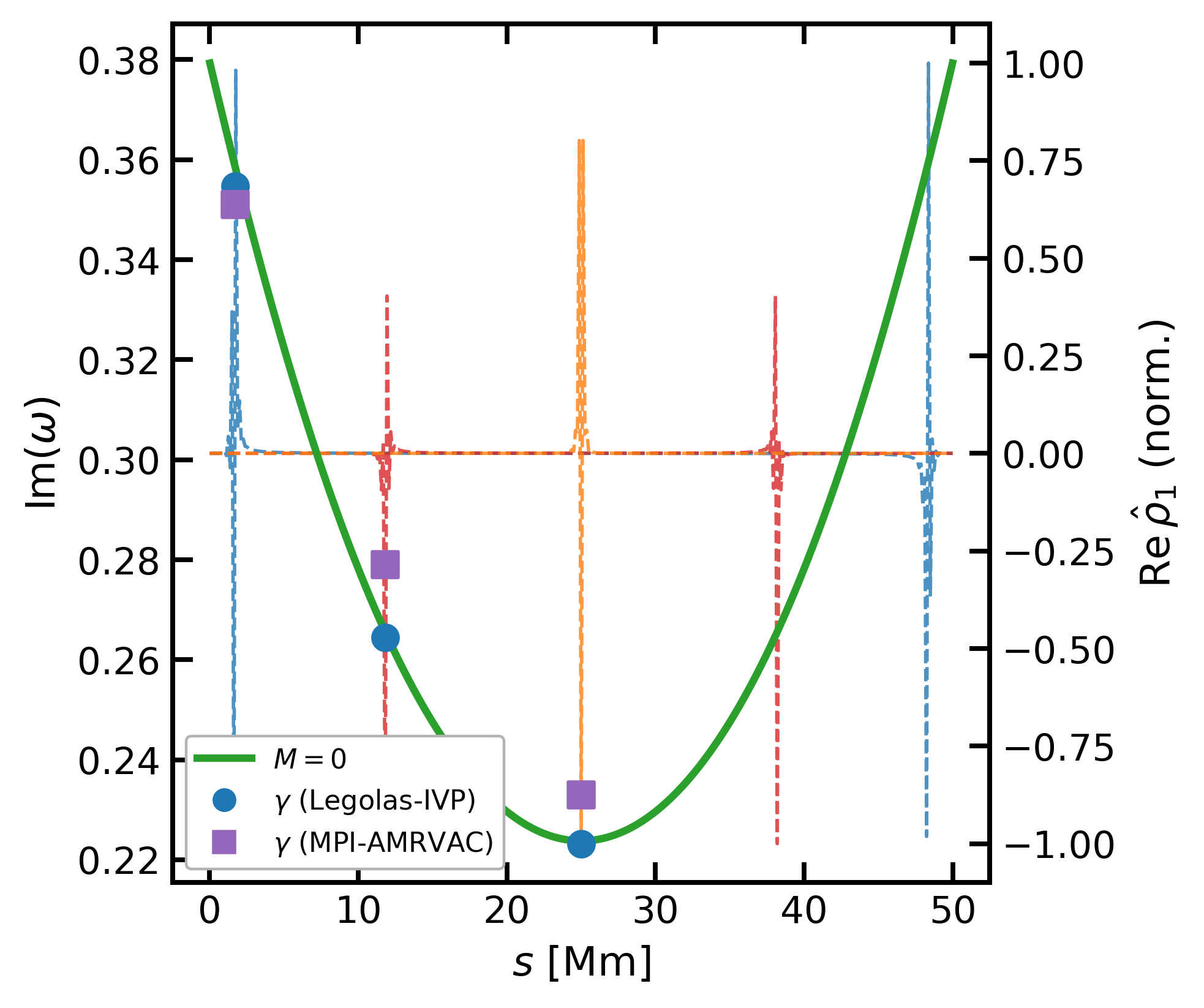}
    \caption{Spatial selectivity of thermal-continuum modes.
    The green curve shows the thermal-continuum branch (M=0) plotted as a function of loop coordinate $s$.
    Coloured dashed curves (right axis) show the real part of three representative thermal-continuum density eigenfunctions.
    Blue circles and purple squares mark growth rates $\gamma$ measured using \textsc{Legolas-IVP} and MPI-AMRVAC respectively.}
    \label{fig:gamma_vs_position}
\end{figure}

To demonstrate this spatial selectivity explicitly, we repeat the isobaric Gaussian experiment with the pulse centred at three different locations along the left-half of the symmetric loop, $s_0 \approx 1.7$~Mm, $11.8$~Mm, and $25$~Mm.
For each pulse location we extract a growth rate using the same procedure.
This time we use a narrower pulse ($\sigma=0.5$~Mm) to better isolate a single localised continuum eigenfunction at each location.
We perform these measurements with both \textsc{Legolas-IVP} and \textsc{MPI-AMRVAC}.
The results are shown in Figure~\ref{fig:gamma_vs_position}, where we display the $M=0$ thermal continuum frequency (green curve) as a function of loop coordinate $s$, together with density eigenfunctions for the three continuum modes (dashed curves) centred at our previously chosen values $s_0$.

The measured growth rates from \textsc{Legolas-IVP} (blue circles) and \textsc{MPI-AMRVAC} (purple squares) follow the same position-dependent trend and lie close to the local continuum growth rates at the imposed pulse centres.
This confirms that distinct segments of the thermal continuum are selectively accessible in initial-value problems, and that, when the perturbation is sufficiently localised, the measured early-time growth rate is primarily controlled by the local continuum eigenvalue associated with the perturbation location.

Returning to the discrete thermal mode (cf. Fig~\ref{fig:loop-spectrum}), we note that this has a smaller growth rate than the thermal-continuum branch at all positions along the loop, so any thermal-instability-triggering perturbation (in entropy) will tend to be dominated at early times by the locally fastest-growing continuum component. 

We point out that the growth rates of the thermal continuum modes are fully determined by the local $M=0$ condition (Eq.~\ref{q-mfac}), which depends only on the analytically prescribed equilibrium quantities at each position along the loop, and is therefore independent of numerical resolution.
This is confirmed in Figure~\ref{fig:gamma_vs_position} by the agreement of continuum growth rates obtained by \textsc{Legolas}, \textsc{Legolas-IVP}, and \textsc{MPI-AMRVAC}, which employ very different numerical resolutions and discretisation schemes.
The spatial profiles of the continuum eigenfunctions are singular: a Frobenius analysis of the governing scalar ODE derived in \citep{keppens_hydrodynamic_2025} shows that the leading coefficient vanishes at $M=0$, giving locally singular solutions with a logarithmic contribution $v_1 \sim \ln|s - s_0|$ near the continuum point $s_0$, analogous to what was shown in the radially-varying MHD cylinder case by \citet{van_der_linden_thermal_1991b}.
Any finite numerical grid approximates this singular behaviour, with spatial localisation sharpening indefinitely with resolution.
A spatial convergence study is therefore not meaningful for these modes.

\subsection{Runaway cooling versus runaway heating}\label{sec:runaway}
While the spectrum contains all information on the linear eigenmodes of the system studied and each of these normal modes is a priori known to be unstable or damped, it yields eigenvalue-eigenvector pairs obeying Eq.~(\ref{q-eig}) that do not specify an arbitrary (complex-valued) amplitude factor for $\mathbf{x}(s)$.
In a time-dependent simulation, we specify an initial perturbation with a clear (real-valued) amplitude and also a sign factor; in practice, we impose either a local temperature deficit or excess by choosing \(T_1(s,0)<0\) or \(T_1(s,0)>0\) in the initial pulse (and the corresponding \(\rho_1\) constraint for the chosen polarisation).
We now perform time-dependent simulations that show how a sign choice in the initial condition can realise either runaway cooling or runaway heating, and we link this with the sign of the initial temperature perturbation.

\begin{figure*}[htb]
    \centering
    \includegraphics[width=1\linewidth]{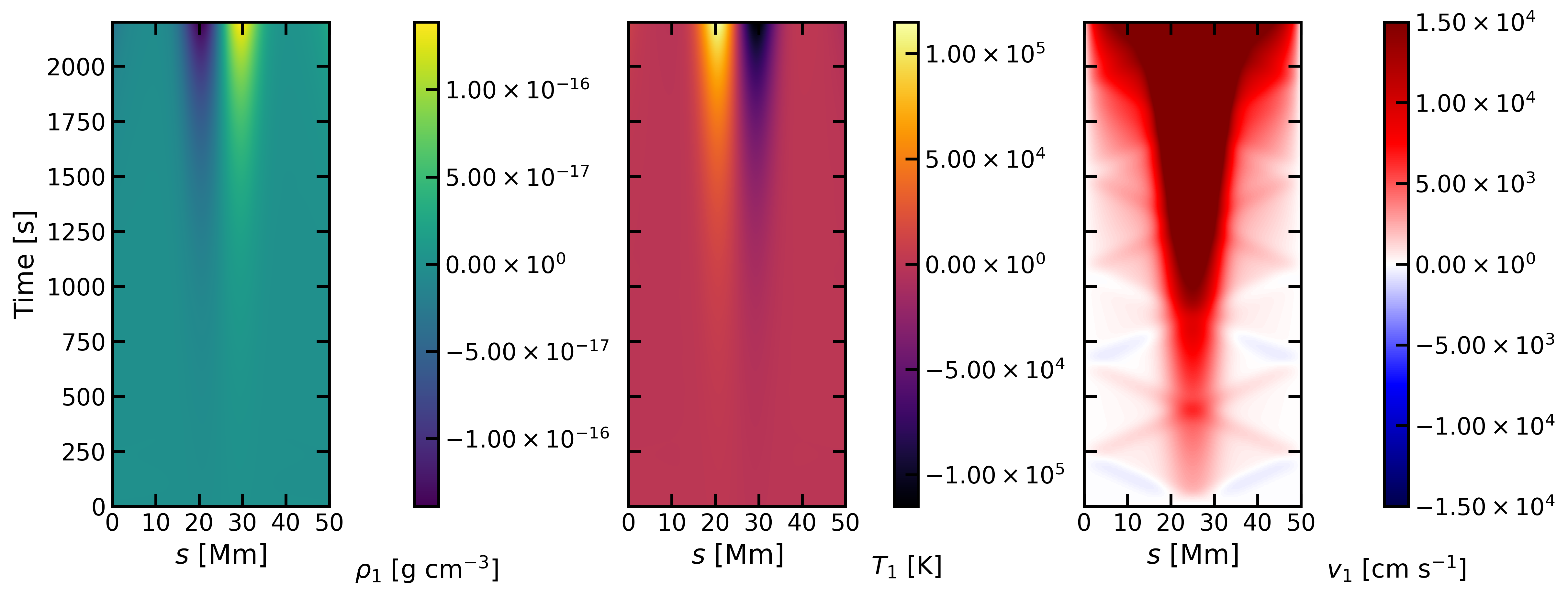}
    \caption{Space--time evolution of a bipolar isobaric perturbation in \textsc{Legolas-IVP}. Perturbation density $\rho_1$ (left), temperature $T_1$ (centre), and velocity $v_1$ (right). 
The positive-temperature pulse (left half) drives runaway heating and rarefaction, while the negative-temperature pulse (right half) drives runaway cooling and condensation, illustrating how the sign of the initial perturbation selects between the two branches of thermal instability (Sect.~\ref{sec:runaway}).}
    \label{fig:heatmap-heating}
\end{figure*}

This is illustrated in Fig.~\ref{fig:heatmap-heating}, where we impose two otherwise identical isobaric Gaussian pulses of opposite sign: a positive temperature perturbation on the left half of the loop and a negative one on the right.
Both perturbations amplify at the same exponential rate, but the left-hand pulse evolves toward runaway heating, while the right-hand pulse evolves toward runaway cooling and condensation.

We note that this symmetric linear picture is a consequence of the linearisation itself: both branches see the same isobaric derivative evaluated at $T_0$, and the linearised system has no knowledge of the global shape of $\Lambda(T)$ away from equilibrium. 
The nonlinear asymmetry between the two branches, and the reason the heating branch reaches a finite plateau while the cooling branch collapses catastrophically, is discussed in Sect.~\ref{sec:nonlinear_transition}.

The sign dependence follows from the linearised net heat-loss term.
We work with the volumetric loss term \(\rho\mathcal{L}\) because that is the quantity that enters the energy equation as a source term.
Linearising around a thermally balanced equilibrium $(\rho\mathcal{L})_0=0$ gives
\begin{equation}
(\rho\mathcal{L})_1=
\left(\frac{\partial(\rho\mathcal{L})}{\partial\rho}\right)_{T}\rho_1+
\left(\frac{\partial(\rho\mathcal{L})}{\partial T}\right)_{\rho}T_1.
\end{equation}
For an approximately isobaric response ($p_1= 0$) the linearised ideal gas law implies $\rho_1 = - \rho_0 \ T_1/T_0$, so that filling into the above yields an expression in terms of the isobaric derivative
\begin{equation}
(\rho\mathcal{L})_1=
\left(\frac{\partial(\rho\mathcal{L})}{\partial T}\right)_{p}T_1.
\label{eq:rhoL_isobaric_paper}
\end{equation}
Using the first law of thermodynamics $\rho T\,dS/dt=-(\rho\mathcal{L})$, 
and the thermodynamic relation between $S_1$ and $T_1$, $S_1=c_pT_1/T_0$ for $p_1=0$, yields
\begin{equation}
\frac{dT_1}{dt}
=-\frac{1}{\rho_0c_p}\left(\frac{\partial(\rho\mathcal{L})}{\partial T}\right)_{p}T_1.
\end{equation}
Here \(c_p=\gamma\mathcal{R}/(\gamma-1)\) is the specific heat at constant pressure, and we use the same entropy variable as before, \(S=p\rho^{-\gamma}\).
Field's isobaric criterion $\left(\partial(\rho\mathcal{L})/\partial T\right)_p<0$ therefore implies exponential growth.
Since $\mathcal{L}>0$ corresponds to net cooling and $\mathcal{L}<0$ to net heating, this shows that $T_1<0$ drives runaway cooling (condensation), whereas $T_1>0$ drives runaway heating.  

This sign dependence does not contradict the second law of thermodynamics. 
The loop plasma is an open system with energy exchange through prescribed heating and optically thin radiative losses. 
During runaway cooling the plasma entropy can decrease locally, but this is accompanied by an increase of entropy in the emitted radiation field (and, more generally, in the combined system consisting of plasma plus its thermal environment). 
The sign of the initial pulse therefore selects whether the perturbation initially increases or decreases the local volumetric loss term \((\rho\mathcal{L})_1\), driving the evolution toward runaway heating or cooling without violating thermodynamic constraints.

\subsection{From linear to nonlinear evolution}
\label{sec:nonlinear_transition}

\begin{figure*}[htb]
    \centering
    \includegraphics[width=1\linewidth]{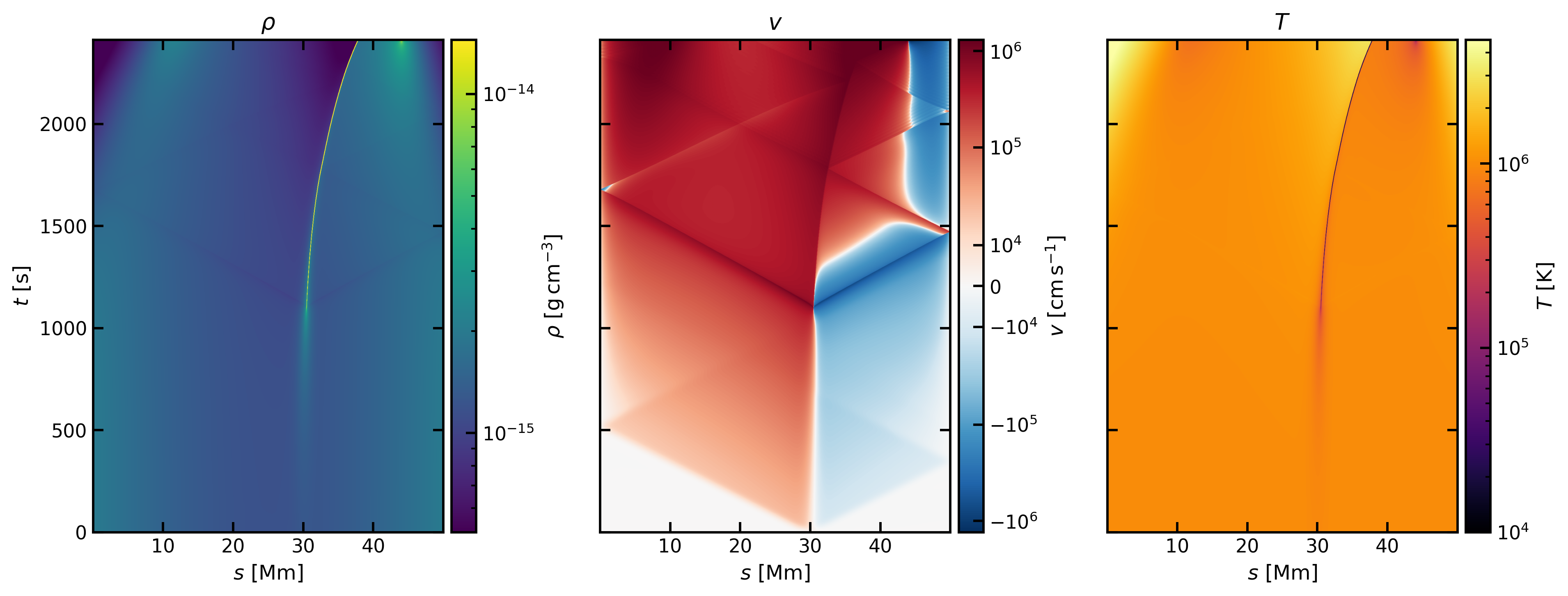}
    \caption{Nonlinear \textsc{MPI-AMRVAC} evolution of the isobaric perturbation ($\sigma=1$~Mm, $s_0=30$~Mm). 
Space--time heatmaps of total density $\rho$ (left), velocity $v$ (centre), and temperature $T$ (right). 
The condensation forms at $s=30$~Mm and slides rightward after $t\simeq1200$~s. Diagonal features in the velocity panel are sub-sonic evacuation waves propagating outward from the condensation site.
Flanking regions of elevated temperature reflect heating of the mass-depleted footpoints under the fixed balanced-heating prescription.}
    \label{fig:amrvac_condensation}
\end{figure*}

\begin{figure}[htb]
    \centering
    \includegraphics[width=1\linewidth]{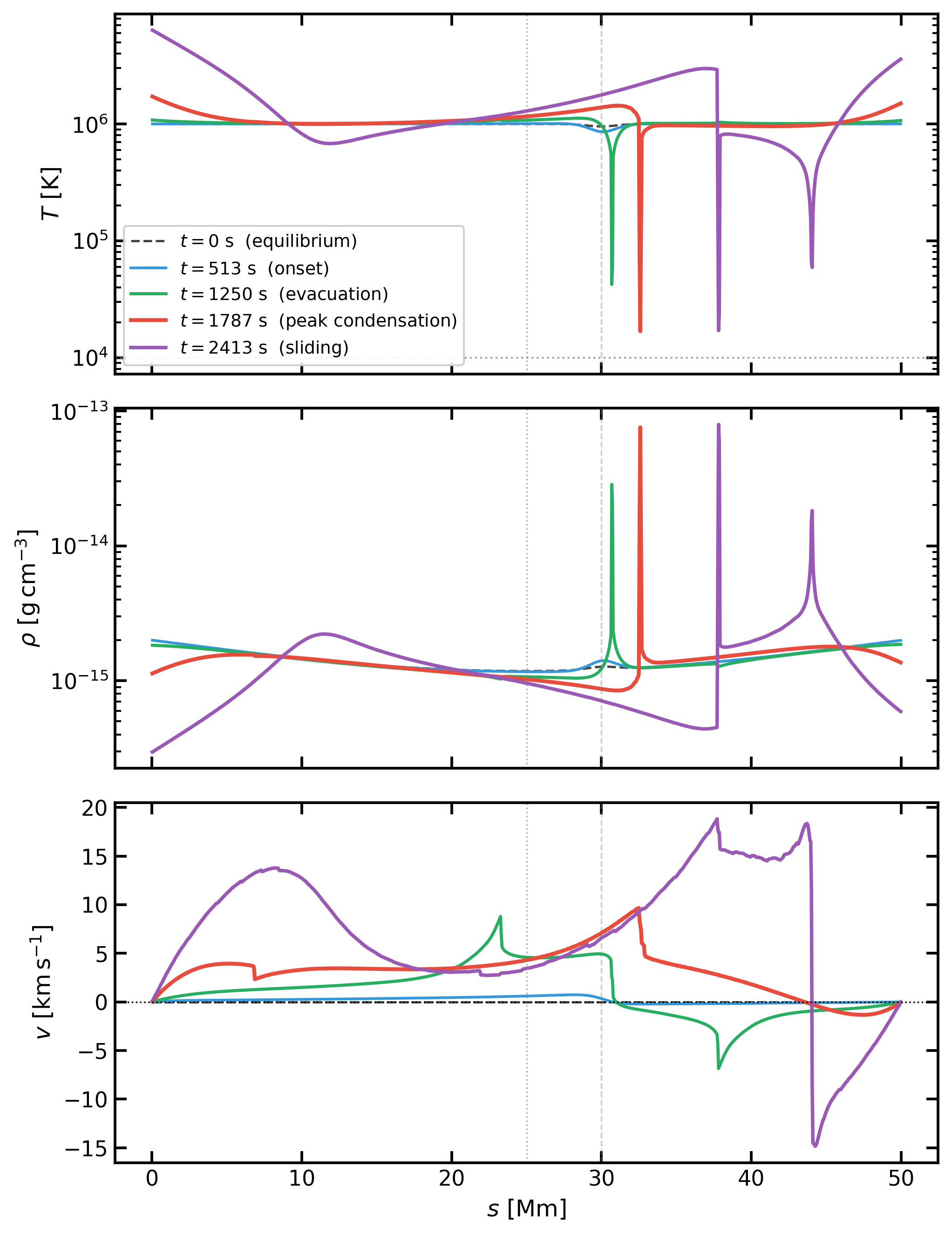}
    \caption{Profiles of temperature $T$ (top), density $\rho$ (middle), and velocity $v$ (bottom) at five representative times. 
The dotted and dashed vertical lines mark the loop apex ($s=25$~Mm) and pulse centre ($s=30$~Mm) respectively. 
The horizontal dotted line in the top panel indicates the temperature floor $T_\mathrm{floor}=10^4$~K.
}
    \label{fig:amrvac_profiles}
\end{figure}

We now use \textsc{MPI-AMRVAC} to follow the instability into the fully nonlinear regime, using an isobaric perturbation centred at $s_0 = 30$~Mm with a narrower width ($\sigma = 1$~Mm) than in Sect.~\ref{sec:growth_rates}, in order to more cleanly isolate the nonlinear development at a single loop location.
The early exponential growth and its agreement with the \textsc{Legolas} spectrum are already discussed there (Fig.~\ref{fig:growth-rates}); here we focus on the subsequent nonlinear development.

Figure~\ref{fig:amrvac_condensation} shows space--time heatmaps of the total density $\rho$, velocity $v$, and temperature $T$ over the full simulation domain and duration.
The condensation initiates at the pulse centre at $s=30$~Mm.
As the instability grows, converging siphon-like flows pull mass from both loop footpoints towards the cooling site, visible as the large-scale positive (negative) velocity on the left (right) half of the loop in the velocity panel.
These flows are already present in the linear stage (Sect.~\ref{sec:time-dep-sim}) and intensify as the density contrast grows.

Figure~\ref{fig:amrvac_profiles} shows profiles of $T$, $\rho$, and $v$ at five representative times: the initial equilibrium, the onset of visible departure from equilibrium ($t\simeq513$~s), evacuation waves ($t\simeq1250$~s), peak condensation ($t\simeq1787$~s), and the sliding phase ($t\simeq2413$~s).
The outward-propagating diagonal features visible in the velocity heatmap at approximately $t\simeq1250$~s are physical waves launched outwards by the collision of the two incoming streams; such `evacuation shocks' are a known feature of coronal prominence/filament formation \citep[e.g.][]{xia_formation_2011}.
In this case, the waves travel at a speed of ${\simeq}50$--$70$~km~s$^{-1}$, sub-sonic relative to the ambient corona ($c_s \simeq 117$~km~s$^{-1}$ at $T=10^6$~K).
These features may therefore be more accurately described as pressure-driven `evacuation waves' rather than fully-developed shocks.
By $t\simeq1500$~s these waves reach the loop footpoints, visible as the abrupt change in the velocity heatmap at the domain boundaries, after which the boundary reflections interact with the ongoing condensation. 
Tests with open (zero-gradient) boundary conditions in place of the perfectly reflecting walls confirmed that these reflections do not significantly affect the subsequent sliding condensation dynamics.

By peak condensation the temperature at the blob has fallen to $T_\mathrm{min}\simeq1.68\times10^4$~K, well into the chromospheric regime, while the local density reaches $\rho_\mathrm{max}\simeq7.55\times10^{-14}$~g~cm$^{-3}$, a factor of ${\sim}60$ above the background value at that location.
This density enhancement is comparable to values reported in other simulations of coronal condensations \citep[e.g.][]{xia_formation_2011, claes_thermal_2020}.

After peak condensation the blob moves rightward under the projected gravitational component $g_\parallel > 0$ for $s > L/2$, as seen in the sliding profile at $t\simeq2413$~s.
The velocity profile at this time shows a coherent positive-velocity region spanning the left half of the loop, with a sharp reversal at the blob location where inflowing material meets.
The secondary density enhancement ahead of the blob at $s\simeq45$~Mm at $t=2413$~s reflects coronal material being swept up and compressed ahead of the sliding condensation.

A notable feature of the nonlinear evolution is the heating of the evacuated coronal regions on either side of the condensation site, visible as the bright orange ($T > 10^6$~K) flanking regions in the temperature heatmap.
Unlike the runaway heating discussed in Sect.~\ref{sec:runaway}, which is driven by the temperature-dependence of the cooling curve, the heating here is primarily driven by the depletion of density in the evacuated regions.
As mass drains from these regions via siphon flows, the local density $\rho$ falls below the equilibrium value $\rho_0$, so  that the radiative losses $\rho^2\Lambda(T)$ decrease while the prescribed background heating $H(s) = \rho_0^2\Lambda(T_0)$ remains frozen at its equilibrium value.
The resulting net heating drives temperature increases in the evacuated regions; this is not a numerical artefact but rather a true realisation of the frozen balanced-heating prescription used.
In a more realistic model, chromospheric evaporation driven by the increased heating would replenish the evacuated regions with dense plasma, quenching the runaway; this is the essence of thermal non-equilibrium cycling \citep{klimchuk_distinction_2019}.
The absence of this feedback in this coronal-only model is a known limitation of the idealised setup, and does not affect the condensation dynamics at the blob itself, which are the primary focus of this study.

The simulation thereby completes the physical picture anticipated by the linear analysis: thermally unstable continuum eigenmodes, selectively excited by the isobaric perturbation, grow exponentially at the rate predicted by \textsc{Legolas}, driving converging siphon flows, and ultimately producing a cool, dense condensation that detaches from close to the apex region and slides toward the footpoint under gravity -- the hallmark nonlinear outcome of thermal instability in a stratified coronal loop.

\begin{figure*}[htb]
    \centering
    \includegraphics[width=1\linewidth]{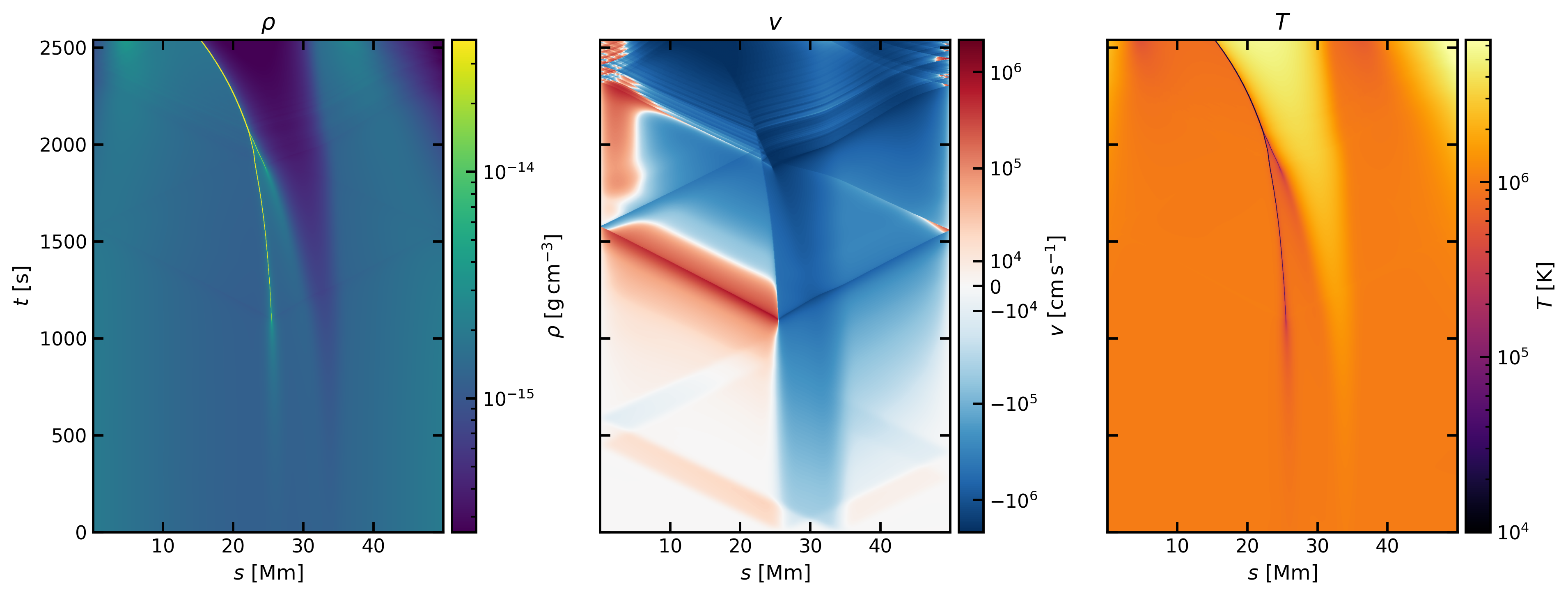}
    \caption{Nonlinear \textsc{MPI-AMRVAC} evolution of the bipolar isobaric perturbation ($\sigma = 1$~Mm). 
    Space--time heatmaps of total density $\rho$ (left), velocity $v$ (centre), and temperature $T$ (right). 
The negative-temperature pulse (right half, $s \gtrsim 25$~Mm) drives condensation and a dense blob that slides toward the left footpoint, while the positive-temperature pulse (left half, $s \lesssim 25$~Mm) drives rarefaction and heating of the evacuated region under the fixed balanced-heating prescription.}
    \label{fig:bipolar_nonlinear_heatmaps}
\end{figure*}

\begin{figure}[htb]
    \centering
    \includegraphics[width=1\linewidth]{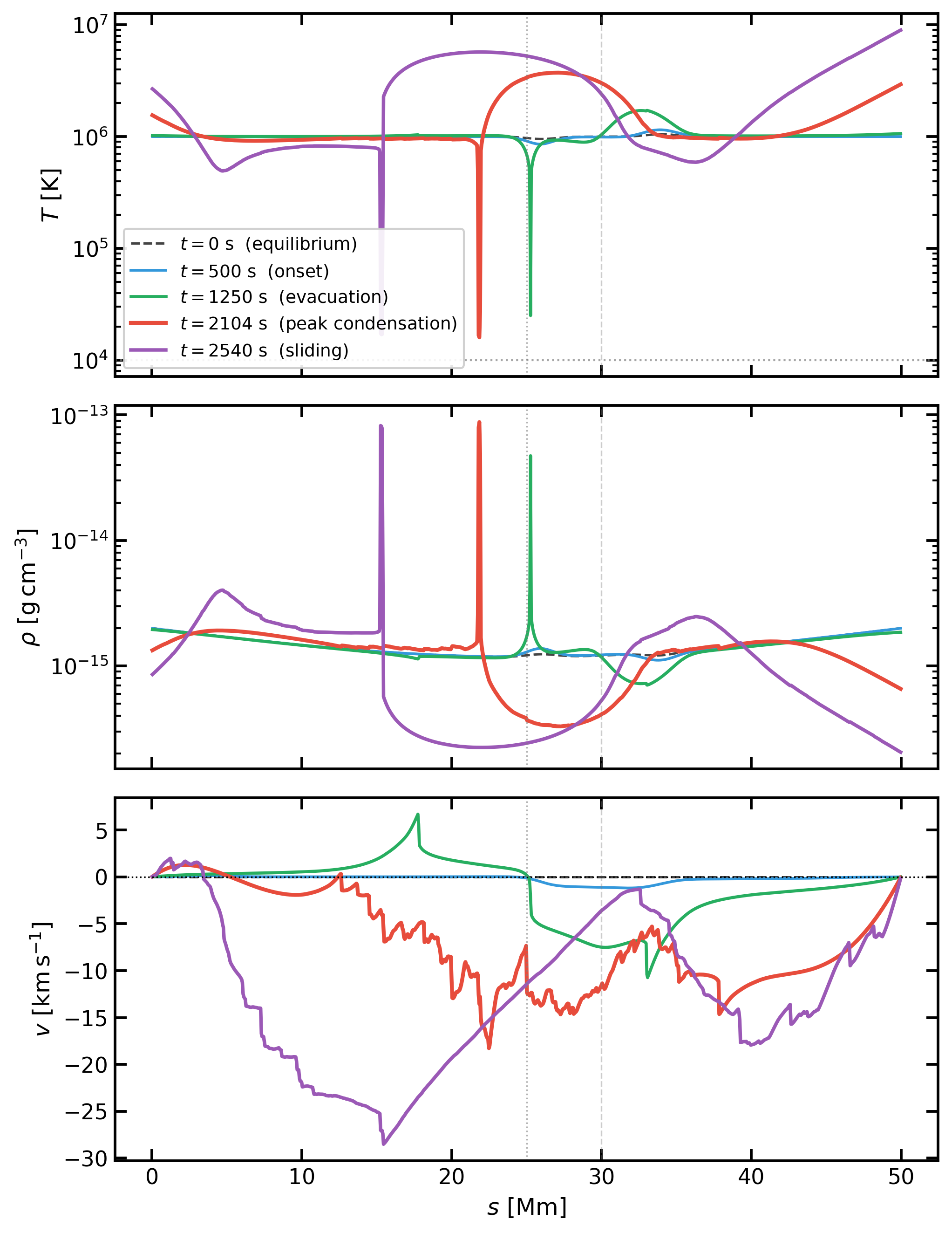}
    \caption{Profiles of temperature $T$ (top), density $\rho$ (middle), and velocity $v$ (bottom) at five representative times for the nonlinear bipolar run. The vertical dotted line marks the loop apex ($s = 25$~Mm). The horizontal dotted line in the top panel indicates the temperature floor $T_\mathrm{floor} = 10^4$~K.
}
    \label{fig:bipolar_nonlinear_profiles}
\end{figure}

The nonlinear evolution also reveals an intrinsic asymmetry between the runaway cooling and heating branches identified in Sect.~\ref{sec:runaway}.
To illustrate this directly, Figs.~\ref{fig:bipolar_nonlinear_heatmaps} and \ref{fig:bipolar_nonlinear_profiles} show the nonlinear \textsc{MPI-AMRVAC} evolution of a bipolar isobaric perturbation analogous to Fig.~\ref{fig:heatmap-heating} but with narrower pulses ($\sigma = 1$~Mm), now followed into the fully nonlinear regime.
The negative-temperature pulse collapses to chromospheric temperatures by $t \simeq 2104$~s and the resulting condensation subsequently slides toward the left footpoint, while the positive-temperature pulse drives a rarefied, superheated region that plateaus near ${\sim}10^7$~K by $t \simeq 2540$~s.
Although both branches are linearly unstable at the same exponential rate, the nonlinear outcome is strongly asymmetric.
This asymmetry is invisible to the linearised system, which approximates $\rho\mathcal{L}$ by its tangent at $T_0$ and therefore predicts identical growth rates for both branches regardless of the global shape of $\Lambda(T)$. 
In the fully nonlinear evolution, however, \textsc{MPI-AMRVAC} evaluates the complete tabulated cooling curve at every timestep. 
Above $T_0$, $\Lambda(T)$ generally decreases with increasing temperature, though with a series of local plateaus associated with emission from different ionisation stages of heavy elements. 
The heating branch is therefore self-reinforcing overall, but a new thermal equilibrium can be reached at any temperature where the curve flattens or turns upward sufficiently for the heating to balance the losses again. 
The nonlinear evolution naturally settles at the first such equilibrium encountered as the temperature rises, rather than diverging catastrophically, in contrast to the cooling branch where no such equilibrium exists until chromospheric temperatures are reached.

\section{Summary and outlook}\label{s-summary}
We presented a unified comparison of spectral, linear initial-value, and fully nonlinear descriptions of thermal instability in a gravitationally stratified 1D hydrodynamic coronal loop subject to optically thin radiative losses and sustained balanced heating.

Using \textsc{Legolas}, we computed the full non-adiabatic spectrum of complex-valued eigenfrequencies $\omega$ and identified acoustic branches together with a thermally unstable branch at $\omega_{\rm R}=0$, including a thermal continuum with strongly localised eigenfunctions.
With the new \textsc{Legolas-IVP} capability, we demonstrated that controlled perturbations selectively excite the expected spectral components: isobaric pulses trigger the unstable thermal response while suppressing acoustic oscillations, whereas isentropic pulses drive weakly damped acoustic behaviour without triggering instability. 
Even in the linear stage, thermal imbalance drives siphon-like flows from the footpoints toward the cooling site. Measured growth rates from \textsc{Legolas-IVP} agree with the spectral predictions and are reproduced independently in nonlinear \textsc{MPI-AMRVAC} simulations of the same equilibrium, establishing a consistent link between eigenvalue spectra and time-dependent evolution.

The nonlinear simulation follows the condensation through runaway cooling to chromospheric temperatures ($T_\mathrm{min} \simeq 1.68 \times 10^4$~K), with a local density enhancement of factor ${\sim}60$ relative to the background. The resulting cool dense blob slides toward the loop footpoint under gravity, consistent with the nonlinear outcome anticipated from the eigenspectrum. 
Mass-depleted flanking regions undergo runaway heating --- a direct realisation of the frozen balanced-heating prescription.

Two limitations of the present study are worth noting. 
Field-aligned thermal conduction is omitted, which is known to have a stabilising effect on thermal instability onset; its effect on the spectral structure is briefly discussed in Appendix~\ref{app:TC}.
The line-tied boundaries also exclude a chromospheric mass reservoir and the evaporation--condensation feedback of thermal non-equilibrium cycling; the heating of the evacuated regions is a direct consequence of this idealisation rather than a numerical artefact.

A natural next step is to extend the model to include thermal conduction and more realistic footpoint physics, enabling the role of conductive stabilisation and chromospheric mass exchange to be quantified within the same framework. 
More broadly, the workflow demonstrated here --- combining spectral analysis, linear IVP simulation, and nonlinear simulation for the same equilibrium --- can be applied to more realistic loop geometries, including multidimensional MHD models.
Incorporating spectral information into 3D loop studies offers a route to interpret condensation formation and coronal rain in terms of the underlying unstable thermal modes and their accessibility.

\begin{acknowledgements}
The computational resources and services used in this work were provided by the VSC (Flemish Supercomputer Center), funded by the Research Foundation Flanders (FWO) and the Flemish Government, department EWI. AK and RK acknowledge funding from the KU Leuven C1 project C16/24/010 UnderRadioSun and the Research Foundation Flanders FWO project G0B9923N Helioskill. JDJ acknowledges funding from the Research Foundation Flanders FWO fellowship 1225625N.
\end{acknowledgements}

\bibliographystyle{aa}  
\bibliography{final.bib} 

\appendix
\section{Verification of the linear initial-value solver}
\label{app:ivp-verification}
The linear initial-value solver used in this work was verified using a wave propagation test in a uniform medium.
We consider a one-dimensional homogeneous background and impose a small-amplitude Gaussian density perturbation.
The system is evolved under an isothermal closure, meaning that the pressure satisfies $p = c_i^2 \rho$ and no energy equation is evolved, where \(c_i\) denotes the isothermal sound speed.
In this limit, the system reduces to the linear wave equation, whose analytical solution is given by d’Alembert's formula,
\begin{equation}
\rho(x,t) = \tfrac{1}{2}\left[\rho(x - c t, 0) + \rho(x + c t, 0)\right].
\end{equation}
Simulations are performed on a domain of length $L=1$ (code units) with sound speed $c_i=0.5$.
Grid resolutions $N_x = \{50, 100, 200, 400\}$ are used, with the time step refined proportionally to maintain a fixed CFL number.
Errors are evaluated at time $t=0.6$, prior to any interaction with the domain boundaries.

Time integration in the solver is carried out using the generalised trapezoidal method with implicitness parameter $\alpha$.
We consider $\alpha=1$ (backward Euler) and $\alpha=0.5$ (implicit midpoint).
For each run, the $L_2$ norm of the error relative to the analytical solution is computed.

\begin{figure}
    \centering
    \includegraphics[width=\columnwidth]{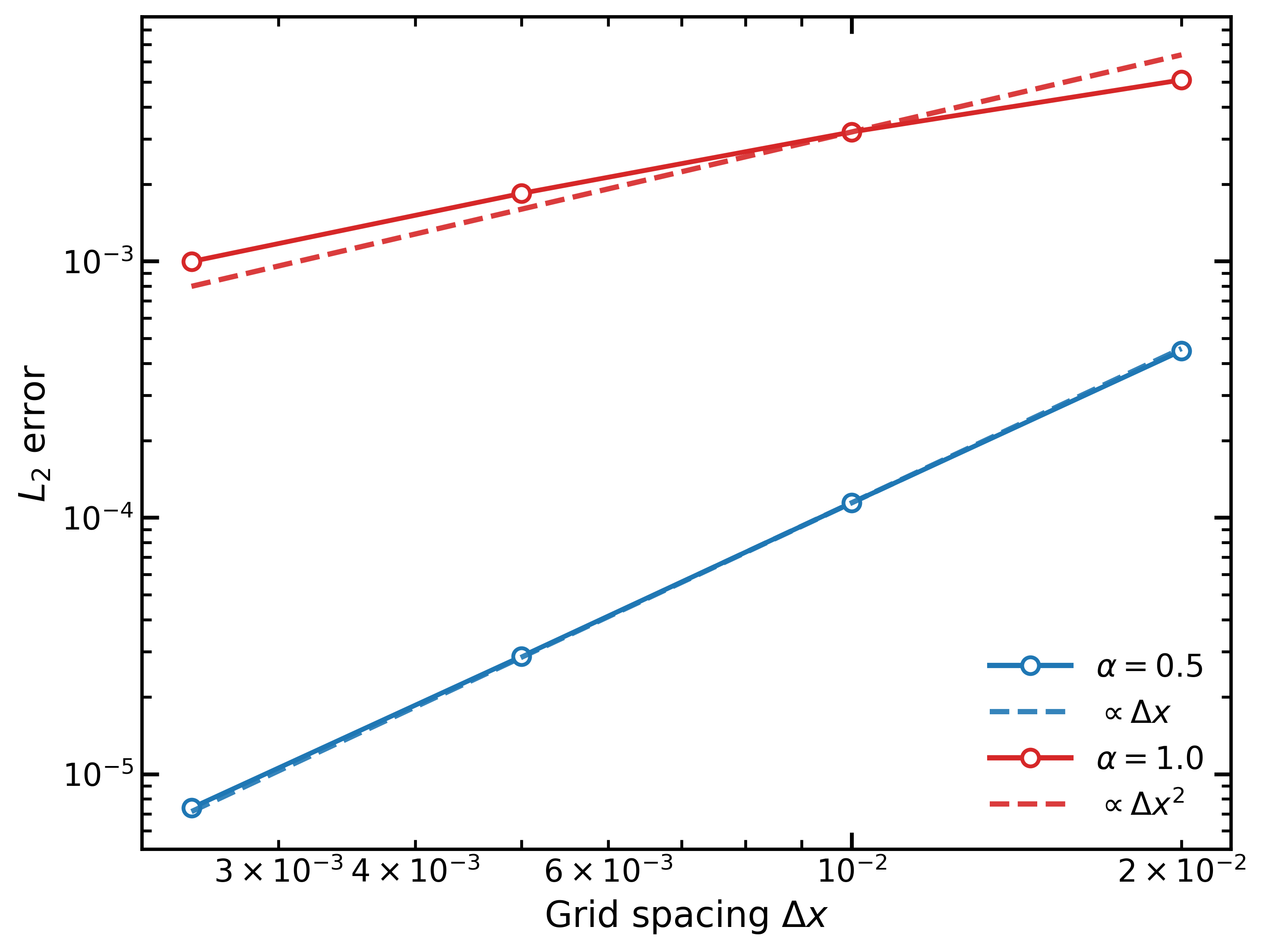}
    \caption{Grid convergence of the linear initial-value solver for wave propagation in an isothermal uniform medium.}
    \label{fig:ivp_convergence}
\end{figure}

Figure~\ref{fig:ivp_convergence} shows the resulting convergence behaviour.
The backward Euler scheme exhibits first-order accuracy, while the implicit midpoint scheme achieves second-order accuracy, in agreement with theoretical expectations.
This confirms that the initial-value solver reproduces the correct linear wave dynamics and is suitable for the analyses presented in the main text.

\section{Effect of field-aligned thermal conduction on the thermal continuum}
\label{app:TC}
\begin{figure}
    \centering
    \includegraphics[width=\columnwidth]{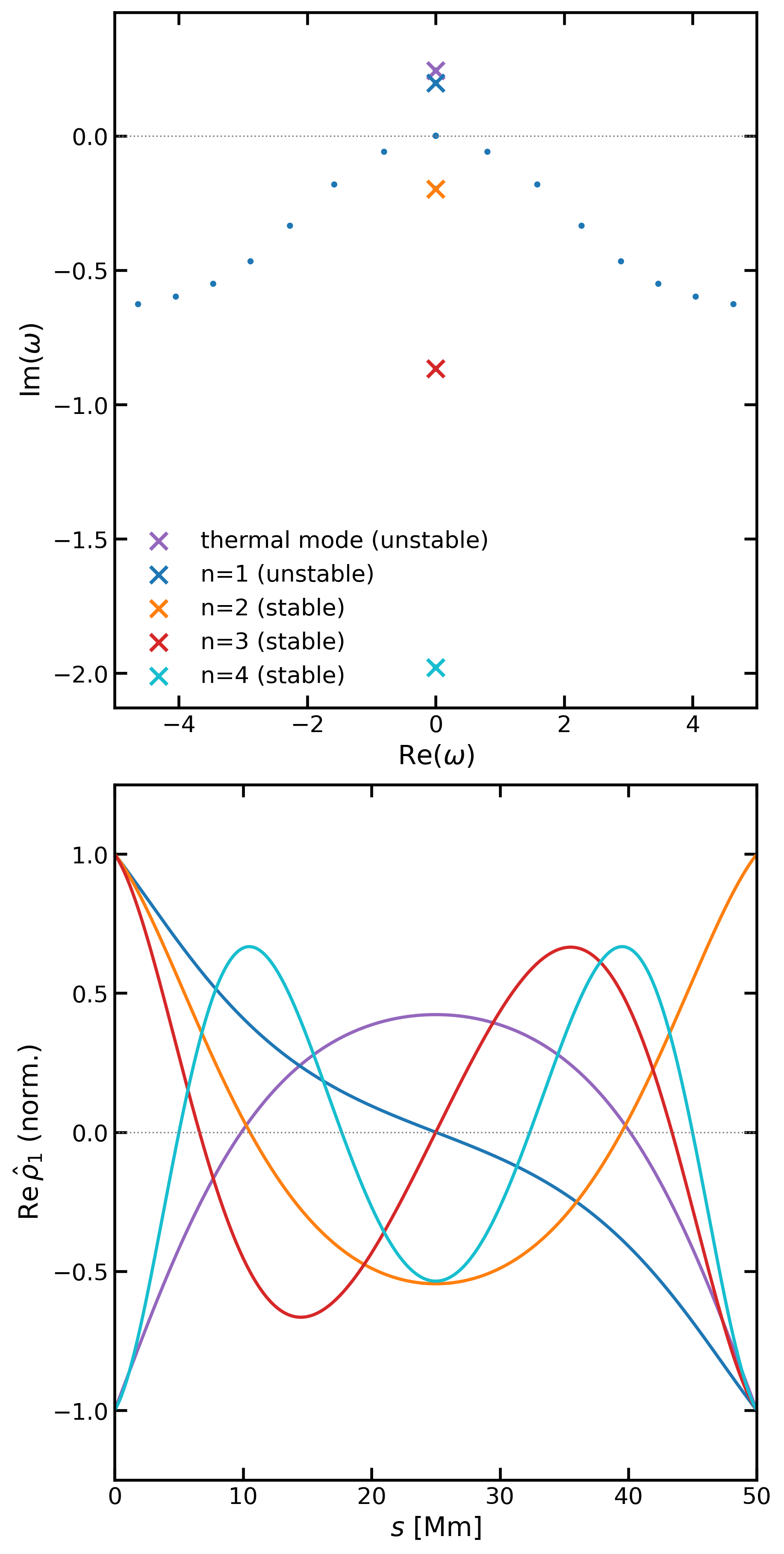}
    \caption{\textsc{Legolas} spectrum for the equilibrium 
    of Sect.~2.2 with Spitzer--H\"arm field-aligned thermal 
    conduction, 
    $N=400$ grid points). \textit{Top:} eigenvalue spectrum; 
    the thermal continuum of Fig.~\ref{fig:loop-spectrum} is 
    replaced by discrete thermal modes (crosses) and the 
    global thermal mode persists (purple marker). Blue dots 
    are acoustic modes. \textit{Bottom:} normalised density 
    eigenfunctions, showing $n$ nodes for the $n$-th mode.}
    \label{fig:TC_spectrum}
\end{figure}

\citet{van_der_linden_thermal_1991b} showed that, for a radially-varying cylindrical MHD equilibrium, perpendicular thermal conduction replaces the thermal continuum with a discrete set of thermal modes.
Figure~\ref{fig:TC_spectrum} shows that an analogous discretisation occurs for the loop equilibrium of Sect.~2.2 when Spitzer--H\"arm field-aligned thermal conduction is included ($\kappa_\parallel\approx0.65$ in code units, $N=400$ grid points), with Neumann (zero heat-flux) boundary conditions imposed on $T_1$ at both footpoints.
The thermal continuum of Fig.~\ref{fig:loop-spectrum} is replaced by a number of discrete thermal modes with smooth, globally extended eigenfunctions displaying $n$ nodes, whose damping rates increase with $n$. 
The discrete global thermal mode of Fig.~\ref{fig:loop-spectrum} persists with an unchanged growth rate but a modified eigenfunction.
The acoustic modes remain largely unaffected.

\end{document}